\documentclass[acmtog]{acmart}
\usepackage{pifont}
\usepackage{array}
\usepackage{tabularx}
\usepackage{multirow}
\usepackage{centernot}
\usepackage{placeins} 

\definecolor{tabmarkok}{HTML}{1B5E20}
\definecolor{tabmarkno}{HTML}{B71C1C}
\definecolor{tabmarkwarn}{HTML}{0D47A1}
\newcommand{\tmarkok}{\textcolor{tabmarkok}{\ding{51}}}
\newcommand{\tmarkno}{\textcolor{tabmarkno}{\ding{55}}}
\newcommand{\tmarkwarn}{%
  \textcolor{tabmarkwarn}{%
    \ding{52}\rotatebox[origin=c]{-9.2}{\kern-0.7em\ding{55}}}}
\AtBeginDocument{%
  }

\setcopyright{acmlicensed}
\copyrightyear{2018}
\acmYear{2018}
\acmDOI{XXXXXXX.XXXXXXX}

\acmJournal{TOG}
\acmVolume{37}
\acmNumber{4}
\acmArticle{111}
\acmMonth{8}

\begin{document}

\title{AnySurf: Any Surface Generation with Directed Edge}


\author{Wenda Shi}
\email{wenda-stone.shi@connect.polyu.hk}
\affiliation{%
  \institution{The Hong Kong Polytechnic University}
  \country{Hong Kong SAR}}

\author{Chenyuan Pan}
\email{tpoto1919@gmail.com}
\affiliation{%
  \institution{PhantomSystem AI}
  \city{Shanghai}
  \country{China}}

\author{Dengming Zhang}
\email{dmz@zju.edu.cn}
\affiliation{%
  \institution{Zhejiang University}
  \city{Hangzhou}
  \country{China}}
  
\author{Yiren Song}
\email{songyiren@sjtu.edu.cn}
\affiliation{%
  \institution{National University of Singapore}
  \country{Singapore}}

\author{Biao Zhang}
\email{biao.z@outlook.com}
\affiliation{%
  \institution{Xi'an Jiaotong University}
  \city{Xi'an}
  \country{China}}

\author{Xingxing Zou}
\email{xingxing.zou@polyu.edu.hk}
\affiliation{%
  \institution{The Hong Kong Polytechnic University}
  \country{Hong Kong SAR}}
  \authornote{Corresponding author}


\begin{abstract}
Open surface components are widely encountered in practical 3D content for real-world industrial pipelines, and they are essential for downstream applications including rendering, physical simulation, and geometric editing. As a representative open surface category, garments have become a natural entry point, and many prior efforts focus on garment generation. These methods have made steady progress with domain-specific representations---e.g., sewing patterns---that split the generation process into 2D pattern prediction followed by physics-based 3D stitching. However, such specialized representation limits scalability and generalization beyond garments, making it difficult to handle broader asset categories such as shoes and accessories. Moreover, most general-purpose 3D generators rely on field-based representations and iso-surfacing, which favor watertight meshes and often fail on open surfaces, producing thin double-layer shells. Trellis2 is a key step forward with a field-free, general 3D representation, yet its open surface outputs still suffer from normal and topology defects.

To address these limitations, we propose \textit{AnySurf}, a unified pipeline for open, closed and hybrid surface 3D assets generation while maintaining faithful face orientation. At its core is \textit{FDG-D}, a directed-edge extension of the Flexible Dual Grid (FDG) representation that preserves critical normal-orientation cues by assigning a direction to each surface-intersecting grid edge. We further introduce a post-training strategy \textit{ROS-FT}, together with a lightweight, plug-and-play \textit{DE-Adapter} that adds only about 1\% more parameters, enabling the model to learn directed edges without compromising its original generative capabilities. Finally, we present \textit{Outfit3D}, a hybrid dataset comprising industry-level open-surface garments and closed-surface accessories. Our approach thus redefines 3D garment modeling, evolving it from a \textit{domain-specific} task into an integral part of the \textit{general} 3D generative landscape. Experiments show that our method outperforms baselines, yielding meshes that are more practical for downstream tasks. We will make our datasets, code, and models publicly available.
\end{abstract}


\keywords{Do, Not, Use, This, Code, Put, the, Correct, Terms, for,
  Your, Paper}

\received{20 February 2007}
\received[revised]{12 March 2009}
\received[accepted]{5 June 2009}

\begin{teaserfigure}
  \includegraphics[width=\textwidth]{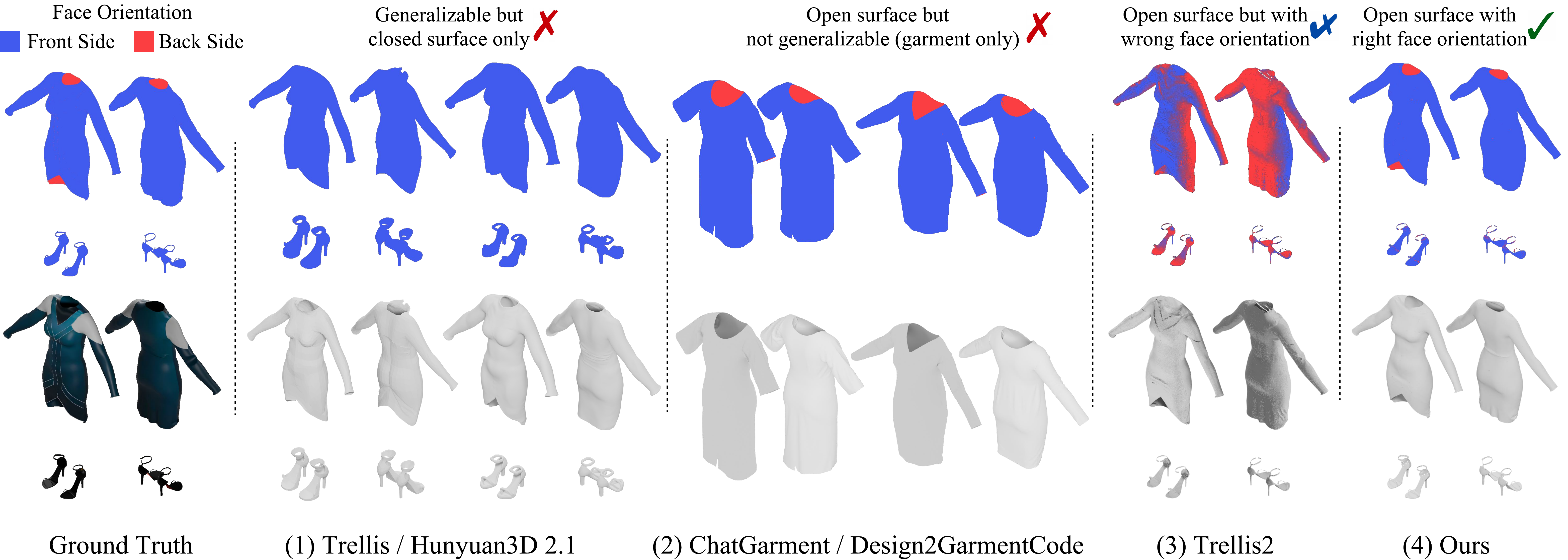}
  \caption{Comparison of methods for generating 3D outfits that combine open surfaces (e.g., garments) and closed surfaces (e.g., shoes and accessories). (1) General-purpose 3D models (Trellis, Hunyuan3D~2.1) are limited to closed (watertight) surfaces. (2) Garment-specific models (ChatGarment, Design2GarmentCode) handle open surface (non-watertight) garment well but do not extend to closed objects such as shoes and accessories. (3) Trellis2 can produce hybrid open-closed geometry but exhibits inconsistent face orientation. (4) Our method yields hybrid geometry with consistent face orientation, most closely matching the ground truth and suitable for downstream applications.}
  \label{fig:teaser}
\end{teaserfigure}

\maketitle

\section{Introduction}

\begin{table}[t]
  \scriptsize
  \begin{tabular}{m{2.4cm}>{\centering\arraybackslash}m{1.5cm}>{\centering\arraybackslash}m{1.7cm}>{\centering\arraybackslash}m{1.5cm}}
  \toprule
  Methods & Closed Surface \newline [General Object] & Open Surface \newline [Thin Garment] & Hybrid Surface \newline [eg, Outfits]\\
  \midrule
  Design2garment {\color{gray}\scriptsize[CVPR 25]} & \tmarkno & \tmarkok & \tmarkno \\
  ChatGarment {\color{gray}\scriptsize[CVPR 25]}  & \tmarkno & \tmarkok & \tmarkno \\
  Dress1to3 {\color{gray}\scriptsize[SIGGRAPH 25]}  & \tmarkno & \tmarkok & \tmarkno \\
  DMap {\color{gray}\scriptsize[SIGGRAPH 25]}  & \tmarkno & \tmarkok & \tmarkno \\
  SewingLDM {\color{gray}\scriptsize[ICCV 25]}  & \tmarkno & \tmarkok & \tmarkno \\
  GarmageNet {\color{gray}\scriptsize[SIG'Aisa 25]} & \tmarkno & \tmarkok & \tmarkno \\
  Trellis {\color{gray}\scriptsize[CVPR 25]} & \tmarkok & \tmarkno & \tmarkno \\
  Hi3DGen {\color{gray}\scriptsize[ICCV 25]}  & \tmarkok & \tmarkno & \tmarkno \\
  Step1X-3D {\color{gray}\scriptsize[ArXiv 25]} & \tmarkok & \tmarkno & \tmarkno \\
  Hunyuan-3D 2.1 {\color{gray}\scriptsize[ArXiv 25]} & \tmarkok & \tmarkno & \tmarkno \\
  Trellis2 {\color{gray}\scriptsize[CVPR 26]} & \tmarkok & \tmarkok & \tmarkok \\
  Ours & \tmarkok + FO & \tmarkok + FO  & \tmarkok + FO \\
  \bottomrule
  \end{tabular}
  \caption{Differences with existing 3D generation methods, classified into thin garment and general object generation. FO means face orientation.}
  \label{tab:rw-1}
  \vspace{-20pt}
\end{table}

Production-grade 3D assets are often \textit{hybrid}, combining closed volumetric surfaces and open surfaces without volumetric enclosure. Accordingly, a practical 3D generation model ought to produce downstream-ready meshes tailored for rendering, physical simulation, geometric editing, and digital manufacturing. However, faithfully modeling open surface components, namely non-watertight geometries, remains a long-standing technical challenge, despite their widespread presence in high-value scenarios. Typical use cases include garments and soft goods (e.g., cloth simulation, virtual try-on, and physical fabrication), foliage and plant structures for large-scale environment reconstruction, as well as thin-walled industrial and precision components such as gaskets, seals, stamped metal parts, and thin protective housings. In particular, garments stand out as the most representative category of open-surface 3D assets.

For thin garment modeling, sewing-pattern based methods are among the most popular: they leverage LLMs/VLMs to synthesize planar patterns (often in SVG-like formats) and stitch them into 3D garments, naturally guaranteeing open surfaces. Some use auto-regressive or diffusion frameworks to either generate vectorized~\cite{he2024dresscode,nakayama2024aipparel,liu2025multimodal,li2025garmentdiffusion3dgarmentsewing} or rasterized~\cite{tatsukawa2025garmentimage} sewing patterns together with edge-wise sewing correspondences and 3D initializations as per-pattern rigid-transformation matrices, or emit high-level parameters~\cite{bian2025chatgarment,guo2025garmentx} and programs~\cite{zhou2025design2garmentcode} of a parametric pattern-making DSL such as GarmentCode~\cite{korosteleva2023garmentcode}. The generated patterns are then draped onto a target avatar using conventional cloth simulation. While this paradigm preserves structural correctness by explicitly producing sewing patterns, it lacks full spatial context and often fails to reproduce fine folds and realistic drape geometry. GarmageNet~\cite{li2025garmagenet} propose a novel garment representation that encodes each pattern as a structured geometry image, effectively bridging the semantic and geometric gap between 2D structural patterns and 3D garment geometries, and with a latent diffusion transformer to synthesize pattern-wise geometry images. However, these \textit{domain-specific} representations struggle to generalize beyond garments. The broader challenge of synthesizing arbitrary 3D assets—particularly combination of open and closed surfaces like full outfits with accessories—remains an open problem.

In parallel, general-purpose 3D generation methods~\cite{chan2021pi, achlioptas2018learning, luo2021diffusion, li2023neuralangelo, gu2021stylenerf, liu2023syncdreamer, long2023wonder3d, liu2024part123} aim to synthesize diverse shapes beyond garments and other domain-specific categories. They build on a range of representations, including point clouds~\cite{achlioptas2018learning, luo2021diffusion}, voxel grids~\cite{smith2017improved, xie2018learning}, and implicit fields such as SDFs~\cite{chen2019learning, autosdf2022}. While point- and voxel-based representations make surface extraction challenging, implicit fields are particularly effective for watertight shapes; supporting open surfaces and complex topologies remains difficult even with improved formulations (e.g., 3PSDF~\cite{chen20223psdf} and TSDF~\cite{sun2021neuralrecon}) or alternative open surface representations (e.g., G-Shell and Surf-D~\cite{liu2024gshell, yu2025surf}). Meanwhile, recent large-scale systems~\cite{zhang2024clay,tochilkin2024triposr,trellis,zhao2025hunyuan3d} explore more scalable latent representations (e.g., VecSet~\cite{zhang20233dshape2vecset} or structured latents~\cite{trellis}) with diffusion transformers, substantially improving generalization---yet the generation of open surface geometry remains a bottleneck.

\begin{figure}[t]
	\centering
	\includegraphics[width=\linewidth]{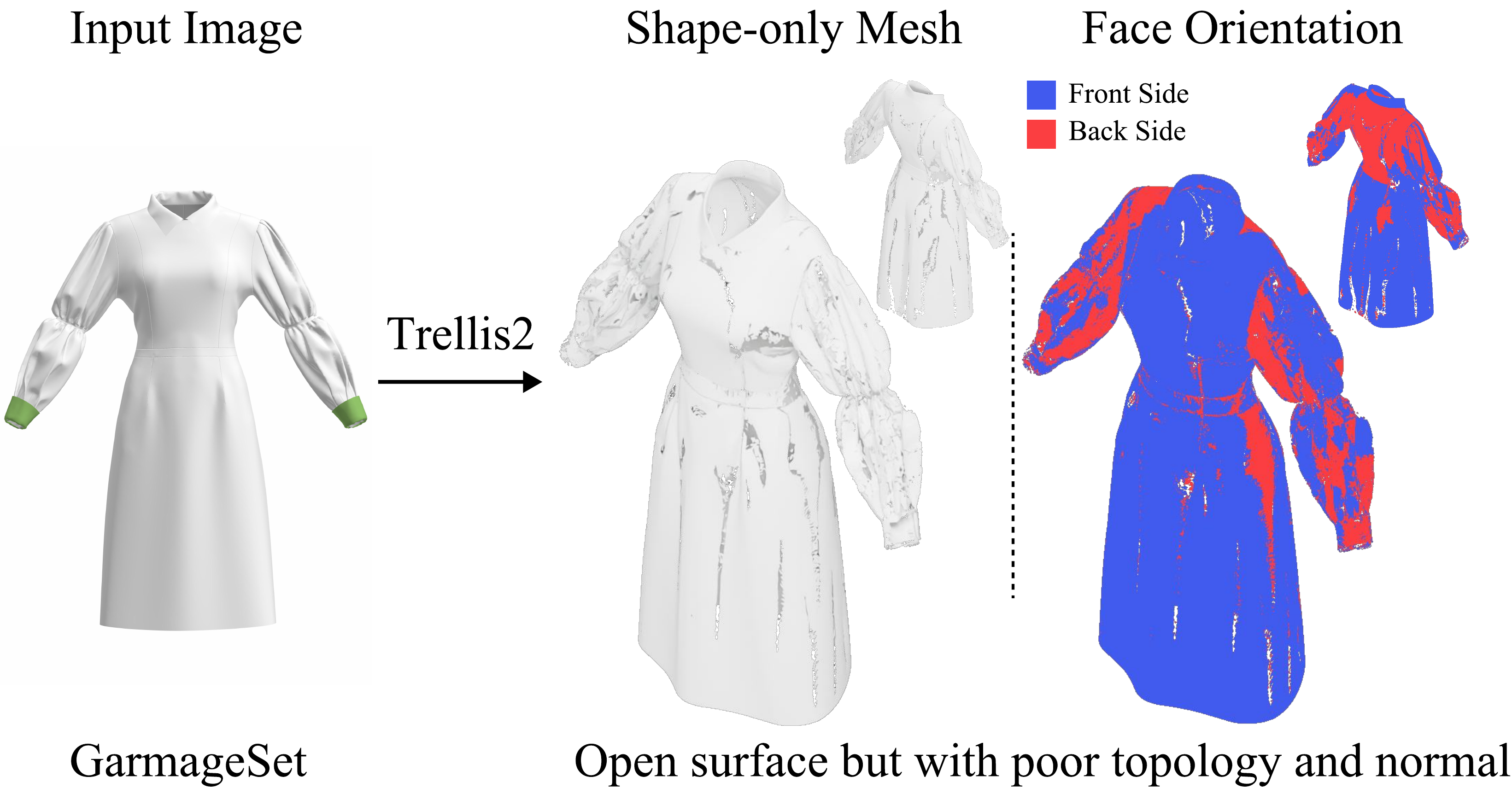}
    \vspace{-18pt}
	\caption{Textureless results of Trellis2 on open surface data (eg, GarmageSet).}
	\label{fig:pre_1_topo_normal}
    \vspace{-10pt}
\end{figure}

This limitation is not merely a modeling gap but also a representation bottleneck. Most general-purpose 3D generative models~\cite{hunyuan3d2025hunyuan3d, trellis, jia2025ultrashape, li2025triposg} ultimately rely on \textit{field-based} shape representations (e.g., SDFs) followed by iso-surfacing (e.g., Marching Cubes/Flexible Cubes~\cite{lorensen1998marching,shen2023flexible}), a pipeline intrinsically biased toward watertight closed surfaces. Geometry without volume is therefore hard to represent faithfully, and open surface structures often collapse into extremely thin yet still closed \textit{double-layer shells}. For garments, such ``thick'' watertight shells introduce multiple nearly coincident layers, making near-contact collision handling ill-conditioned and prone to interpenetration.

To solve this problem, Trellis2~\cite{trellis2} marks a key turning point by introducing a \textit{field-free} and \textit{general} representation, breaking the long-standing dependence on fields and iso-surfacing while retaining generality. However, as shown in Fig.~\ref{fig:pre_1_topo_normal}, its generated open surface garment suffer from inconsistent face orientation and topological defects (e.g., non-manifoldness and local connectivity issues). This reveals a critical issue: even when a mesh is nominally open, inconsistent normals can be catastrophic. They not only break the appearance (shading and material response) but also severely destabilize downstream physics-based simulation and geometry processing. Therefore, our goal is to enable \textit{general} 3D generation that supports open surfaces with reliable face orientation, without sacrificing the inherent generative capabilities of base models.

To achieve this goal, we propose a new 3D representation \textit{FDG-D} (Flexible Dual Grid with Directed Edge) that preserves normal information during dualization. We also propose a post-training strategy, \textit{ROS-FT} (Real Open Surface Fine-Tuning), with a plug-and-play \textit{DE-Adapter} (Directed Edge Adapter) to learn directed edges while keeping the base model's capabilities. Finally, we introduce a hybrid dataset, \textit{Outfit3D}, to evaluate mixed open/closed assets rather than only pure open surfaces. Our contributions are summarized:
\begin{itemize}
  \item \textit{AnySurf} framework. A unified 3D generation pipeline capable of synthesizing any 3D surface (open, closed, or hybrid) with correct face orientation. This establishes a new paradigm for garment modeling, shifting the field from domain-specific workflows toward universal 3D generative pipelines.
  \item \textit{FDG-D} representation. A minimal yet effective 3D representation that explicitly encodes directed edges, enabling robust reconstruction and generation of arbitrary surfaces while preserving consistent face orientations.
  \item \textit{ROS-FT} strategy \& \textit{DE-Adapter}. A progressive post-training strategy equipped with a lightweight module (adding only $1.16\%$ parameters) to restore face orientation without compromising the powerful generative priors of the base model.
  \item \textit{Outfit3D} dataset. The first comprehensive hybrid dataset containing both open surfaces (thin garments) and closed surfaces (accessories), featuring industry-level high-precision outfits with textures and UV (2D patterns) information.
  \item Extensive Evaluation. We conduct experiments and ablation studies against existing baselines, demonstrating the superiority and effectiveness of our proposed method across diverse topological domains.
\end{itemize}

\section{Related Works}
\subsection{Garment Generation}
Garments are a representative and practically important domain for open surface generation. Existing approaches can be organized into three paradigms: \textit{pattern-based} methods that explicitly model 2D sewing patterns and seams, \textit{geometry-based} methods that directly synthesize 3D garment surfaces, and hybrid \textit{pattern+geometry} methods that jointly generate both patterns and surface initialization.

\textit{Pattern-based} pipelines typically proceed in two stages: they first predict 2D patterns and seam connectivity as sewing patterns, and then stitch patterns along seams and drape them on a target avatar with cloth simulation. Depending on the 2D pattern representation, prior work falls into three families.
(i) Vector-quantization (VQ) methods cast pattern prediction as sequence modeling over discrete 1D tokens. NeuralTailor~\cite{korosteleva2022neuraltailor} reconstructs patterns from unstructured point clouds with an LSTM decoder and point-level attention; later work expands inputs (text in DressCode~\cite{he2024dresscode}, images in SewFormer~\cite{liu2023sewformer}) and couples prediction with differentiable simulation so that simulated drape matches the image~\cite{li2025dress}. Diffusion-based~\cite{liu2023sewformer,li2025garmentdiffusion3dgarmentsewing} and LLM-based~\cite{nakayama2024aipparel} variants further scale the VQ formulation to larger, more complex GarmentCodeData and more modalities.
(ii) Template-based methods address structural validity (e.g., seam correctness) by targeting GarmentCode~\cite{korosteleva2023garmentcode} and generating high-level parameters or DSL programs~\cite{bian2025chatgarment,zhou2025design2garmentcode,guo2025garmentx} to enforce constraints and improve pattern quality.(iii) Image-based encodings preserve the 2D layout of patterns via explicit~\cite{chen2022structure,tatsukawa2025garmentimage} or implicit~\cite{li2024isp} raster representations. The extra capacity supports UV-aligned maps of 3D drape~\cite{li2024garment,li2024reconstruction,li2025single} and pairs naturally with modern image generative models~\cite{gu2002geometry,yan2024object,yu2024super}, at the cost of redundancy.

\textit{Geometry-based} methods skip explicit patterns and model garment geometry directly in 3D. Implicit approaches use unsigned distance fields (UDF)~\cite{yu2025surf,chen2024neural}, manifold distance fields~\cite{liu2024gshell}, or Gaussian splatting~\cite{liu2024clothedreamer, rong2025gaussian}, often with diffusion or GAN-based generators for plausible assets and dynamics~\cite{xie2024physgaussian,rong2025gaussian}. Extracting usable triangle or quad meshes from these representations still leans on iso-surface extraction, which is fragile for thin sheets with open boundaries. Another family deforms a fixed garment template and registers it to an avatar via optimization~\cite{zhu2022registering,qiu2023rec} or differentiable simulation~\cite{li2024diffavatar,sarafianos2024garment3dgen}; although the topology stays controlled, but expressiveness is limited by the template itself.

\textit{Dual-stream (Pattern+geometry)} methods aim to jointly generate patterns, sewing seam, and a 3D initialization compatible with physics-based simulation. For example, \cite{li2025garmagenet} proposes a unified framework that automates 2D pattern creation, seam construction, and 3D garment initialization, following~\cite{yan2025object}.

Overall, although these methods are able to generate open surfaces (thin garments), but garment-specific pipelines rely on domain-tailored structure (pattern--seam programs, pattern tokenizations, or fixed garment templates), which restricts scalable training data and can hurt generalization relative to more general 3D generators.

\subsection{General Object Generation}

General-purpose 3D generation has witnessed a paradigm shift, evolving through distinct representational and architectural stages to synthesize diverse, high-fidelity assets beyond domain-specific categories. This evolution can be categorized into three stages.

\textit{Stage 1: Early Generative Models.} Initially, 3D generation methods~\cite{chan2021pi, achlioptas2018learning, luo2021diffusion, li2023neuralangelo, gu2021stylenerf, liu2023syncdreamer, long2023wonder3d, liu2024part123} predominantly relied on explicit, discrete representations, such as point clouds~\cite{achlioptas2018learning, luo2021diffusion} and voxel grids~\cite{smith2017improved, xie2018learning}. While intuitive, these structures often struggle with high-resolution surface extraction. To address this, implicit fields---particularly Signed Distance Functions (SDFs)~\cite{chen2019learning, autosdf2022}---emerged as the standard paradigm. Implicit fields are exceptionally effective for ensuring continuous, watertight geometries, demonstrating unprecedented quality in closed surface modeling.

\textit{Stage 2: Shift to Diffusion Transformers.} As 2D image synthesis pivoted toward scalable architectures, 3D generation similarly adopted Diffusion Transformers (DiTs) as a milestone breakthrough. Moving beyond localized U-Net structures, pioneering works such as DiT-3D explored plain transformers for high-fidelity shape generation by directly denoising voxelized point clouds, demonstrating superior scalability and structural coherence. This transition enabled the integration of structured latent spaces (e.g., VecSet~\cite{zhang20233dshape2vecset} or sparse voxels~\cite{trellis}) capable of decoding into multiple output formats, including radiance fields, 3D Gaussians, and explicit meshes~\cite{liu2024part123, tochilkin2024triposr}, serving as a fundamental paradigm shift in 3D generation.

\textit{Stage 3: Large-Scale 3D Foundation Models.} This architectural evolution has unfolded alongside the rapid expansion of massive 3D asset datasets (e.g., Objaverse~\cite{deitke2023objaverse, deitke2024objaverse} and TexVerse~\cite{zhang2025texverse}). Benefiting from the transformer architecture's scalability, recent large-scale systems---such as Trellis~\cite{trellis}, Hunyuan3D~\cite{hunyuan3d2025hunyuan3d, zhao2025hunyuan3d}, CLAY~\cite{zhang2024clay}, and others~\cite{chen20243dtopia, yang2025pandora3d, li2025step1x, li2025triposg}---leverage billion-parameter backbones to achieve general 3D asset creation with high-quality shapes and textures, improving generalization across diverse categories.

Despite these remarkable advancements, the downstream usability of the generated geometry remains a fundamental bottleneck. Specifically, the prevailing reliance on SDFs or volumetric latents inherently assumes watertight, manifold surfaces. Consequently, even with improved formulations (e.g., 3PSDF~\cite{chen20223psdf} and TSDF~\cite{sun2021neuralrecon}), these state-of-the-art methods systematically struggle to accurately reconstruct and generate thin-shell structures, open surfaces, and complex self-intersecting topologies. While alternative open surface representations---such as G-Shell~\cite{liu2024gshell} and Surf-D~\cite{yu2025surf}---have been proposed to successfully handle arbitrary topologies including thin garments, their complex formulations (e.g., manifold distance fields or Unsigned Distance Fields) severely limit their scalability. They are computationally intensive and difficult to seamlessly integrate into the massive transformer backbones that drive modern 3D foundation models. This underscores the necessity for new representational paradigms tailored for open surface 3D generation that are both topologically expressive and highly scalable.

\subsection{3D Representations For Open Surface}
Neural implicit functions are a widely used family of 3D representations due to their ability to model surfaces with arbitrary topology. They map query coordinates to occupancy values~\cite{mescheder2019occupancy} or to signed/unsigned distance fields~\cite{park2019deepsdf, chibane2020neural}, and can be learned from diverse 2D/3D supervision signals~\cite{wang2021neus, ma2021npull, chen2024inferring, takeshi2024multipull, li2022learning}. To better handle open surfaces, which lack a well-defined inside--outside notion, many methods adopt Unsigned Distance Functions (UDFs) and predict the unsigned distance from arbitrary points to the surface~\cite{chibane2020neural, guillard2022meshudf, long2023neuraludf, zhou2024cap, zhang2024vrprior}; recent work further couples UDF learning with 3D Gaussian Splatting for image-based reconstruction, e.g., GaussianUDF~\cite{li2025gaussianudf}. However, these pipelines typically require a separate surface extraction stage---often adapting marching cubes~\cite{hou2023robust, kobbelt2001feature, lorensen1998marching} or dual contouring~\cite{ju2002dual, chen2022neural}---and can be sensitive to sampling and optimization choices, leading to artifacts such as holes, noise, or unstable orientation. Moreover, most prior open surface reconstruction methods are primarily developed for single-object reconstruction training on relatively small 3D object datasets, making it challenging to scale up to general 3D asset generation.

In contrast, recent large-scale models like Trellis2~\cite{trellis2} leverage native, compact structured latents---specifically, O-Voxel representations featuring a Flexible Dual Grid (FDG) for shape and Volumetric Surface Attributes (VSA) for textures. This allows them to directly represent open surfaces, non-manifold geometry, and internal structures without relying on costly iso-surface extraction, marking a significant step toward versatile, open surface 3D generation. However, the standard FDG formulation inherently discards explicit normal information during the grid construction process. Consequently, the generative model struggles with chaotic and inconsistent face orientations, severely limiting the usability of the generated assets in normal-sensitive downstream applications, such as physics-based simulation and high-quality rendering.

\begin{figure*}[t]
	\centering
	\includegraphics[width=\linewidth]{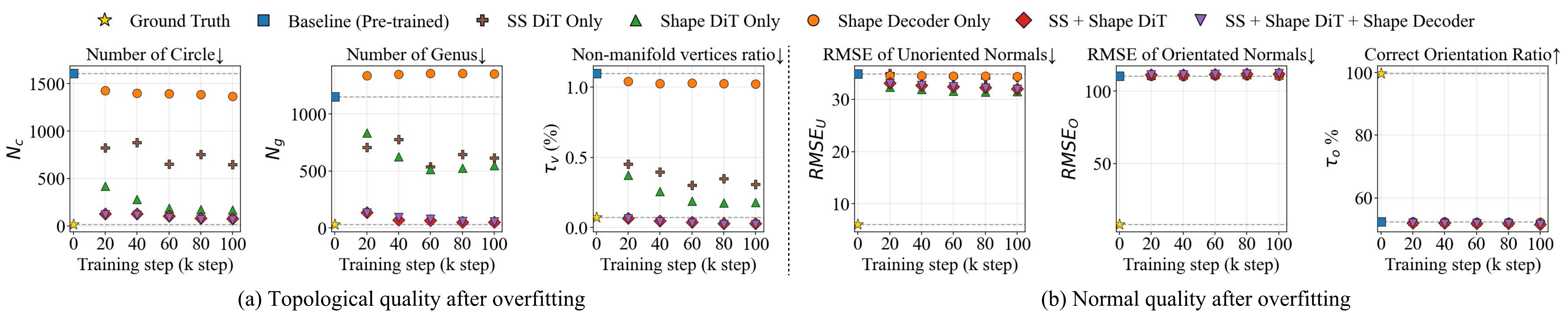}
    \vspace{-18pt}
	\caption{The topological and normal quality with different training steps of Trellis2 on open surface data. SS DiT is the Sparse Structure DiT.}
	\label{fig:finding1}
    \vspace{-10pt}
\end{figure*}
 
\section{Motivation}
\subsection{Preliminaries}
Prior 3D generative models~\cite{trellis, yang2024hunyuan3d, zhao2025hunyuan3d, li2025triposg} can synthesize complex, high-fidelity meshes, yet most rely on signed distance fields (SDFs) and marching cubes that inherently favor closed, watertight surfaces and thus struggle with open surface geometry. To achieve open surfaces, Trellis2~\cite{trellis2} introduces a field-free 3D representation \textit{O-Voxel}, consisting of a shape component, \textit{Flexible Dual Grid} (FDG), and a material component, \textit{Volumetric Surface Attributes} (VSA). FDG constructs a dual grid over a regular voxel lattice: it assigns one dual vertex to each active voxel and activates a quadrilateral dual face (quad for short) for each voxel edge intersected by the input mesh, thereby connecting adjacent dual vertices. Inspired by Dual Contouring~\cite{ju2002dual,chen2022neural} but without any underlying scalar field (e.g., an SDF), FDG directly computes mesh--edge intersections and uses the resulting points and normals as Hermite data \(\{\boldsymbol{q}_i,\boldsymbol{n}_i\}\). For each active voxel, the dual vertex \(\boldsymbol{v}\) is obtained by minimizing the quadratic error function (QEF):
\begin{equation}\label{eq:qef}
\min_{\boldsymbol{v}\in\text{voxel}} e(\boldsymbol{v}) =
\sum_i d_{\Pi,i}^2
+ \lambda_{\text{bound}} \sum_j d_{L,j}^2
+ \lambda_{\text{reg}}\, d_{\hat{\boldsymbol{q}}}^2.
\end{equation}
Compared with the original Dual Contouring QEF term \(d_{\Pi,i}^2=(\boldsymbol{n}_i\!\cdot\!(\boldsymbol{v}-\boldsymbol{q}_i))^2\), FDG adds a boundary-alignment term \(d_{L,j}^2\) and a regularizer \(d_{\hat{\boldsymbol{q}}}^2=\|\boldsymbol{v}-\bar{\boldsymbol{q}}\|^2\), improving open surface fidelity while stabilizing the optimization. Given the solved dual vertices and edge intersection flags, conversion is efficient and bidirectional: mesh\(\to\)O-Voxel (FDG) computes Hermite data and solves Eq.~\eqref{eq:qef}, whereas O-Voxel (FDG)\(\to\)mesh connects neighboring dual vertices across intersected edges to form quads (then split into triangles).

Beyond the representation, Trellis2 is a large-scale, three-stage 3D-native DiT pipeline pre-trained on 970K 3D assets: sparse structure generation, shape generation, and material generation, built upon VAE and flow-matching DiT paradigm~\cite{trellis, esser2024scaling}. To better understand Trellis2 on open surfaces, we conduct pre-experiments to investigate the following research questions:
{\renewcommand\labelenumi{\arabic{enumi})}
\begin{enumerate}
\item \textbf{RQ1:} the boundary of Trellis2 in open surface generation;
\item \textbf{RQ2:} the underlying cause of persistent normal issues.
\end{enumerate}}

\subsection{What's Upper Bound on Open Surface Generation?}

To establish Trellis2's upper bound for open surface generation, we deliberately overfit its shape-related components (sparse structure DiT, shape DiT, and shape VAE decoder) on a 500-instance subset of open-surface garments (GarmageSet~\cite{li2025garmagenet}). By factoring out generalization, this setup isolates the model's inherent capacity ceiling. As shown in Fig.~\ref{fig:finding1}, the model clearly learns to represent open surfaces: jointly training all shape components drastically reduces topological errors, with the number of circles ($N_c$) dropping from over 1500 to roughly 100. In contrast, normal-related metrics show zero improvement throughout training. The correct face orientation ratio ($\tau_o$) remains persistently at $\sim50\%$, while the oriented normal RMSE ($\mathrm{RMSE}_O$) hovers $\sim110^\circ$.

\noindent\textbf{Finding 1.} Even under intentional overfitting, the normal metrics (especially orientation) remain far from the ground truth, which is substantially more so than the topology metrics, motivating our next question on why normal issues persist?

\subsection{Why are Normal Issues Persistent?}

\begin{figure}[t]
	\centering
	\includegraphics[width=\linewidth]{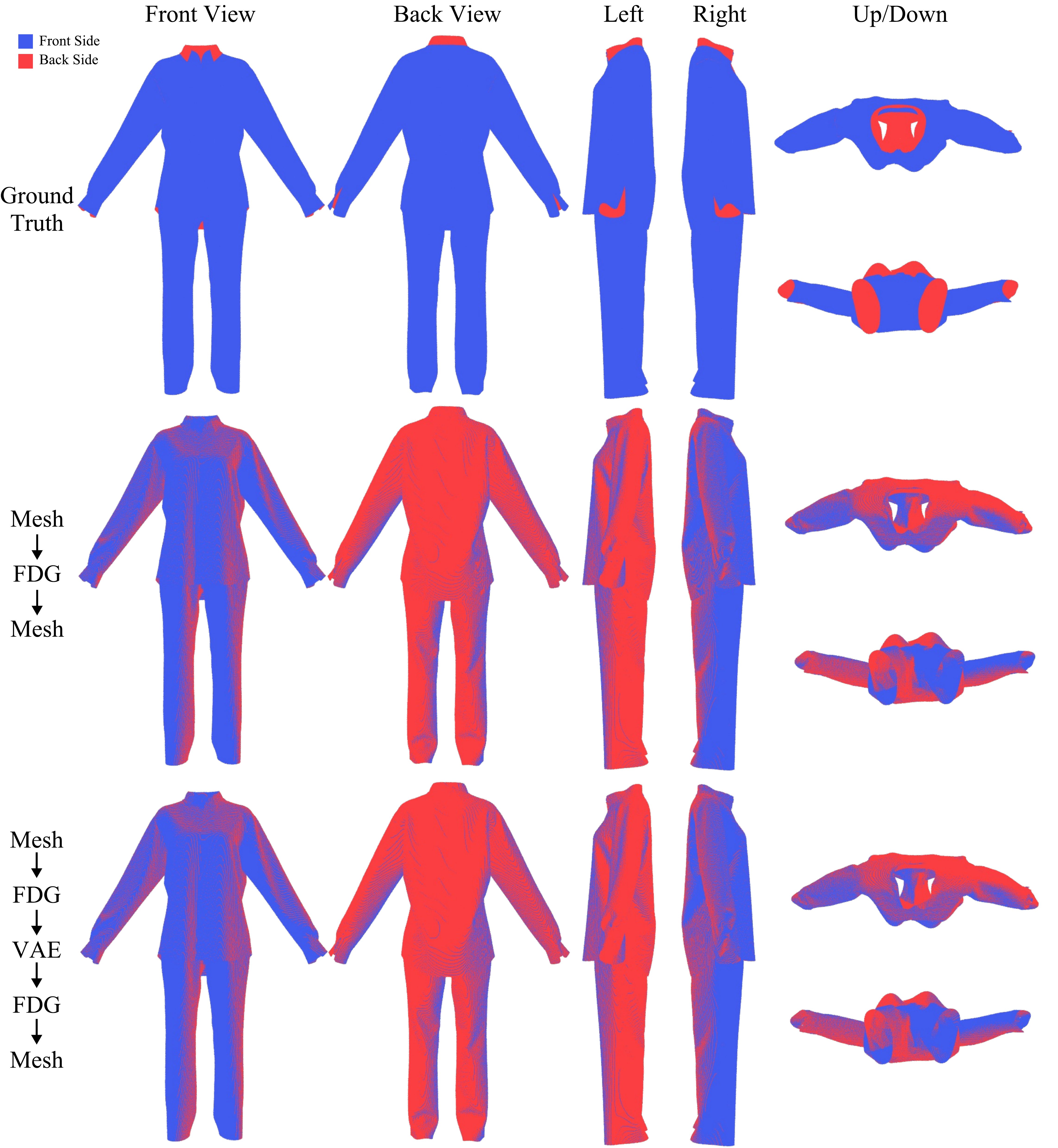}
  \vspace{-18pt}
	\caption{Incorrect face orientation already emerges in the Flexible Dual Grid (FDG) representation and persists after Shape-VAE reconstruction.}
	\label{fig:finding2}
\end{figure}

The generation quality of DiT is fundamentally bounded by the reconstruction quality of its preceding VAE~\cite{yao2025reconstruction, zheng2025diffusion}, which in turn is constrained by whether the 3D representation can faithfully encode the original mesh. Because the DiT is trained to model the latent distribution of the VAE, and the VAE is trained to reconstruct the intermediate 3D representation rather than the raw mesh, any geometric or topological inaccuracies introduced at the representation or VAE stage will inevitably persist and even be amplified during generation.

Motivated by this observation, we skip the DiT stage and focus on the 3D representation and the Shape-VAE reconstruction process. Fig.~\ref{fig:finding2} visualizes the failure mode: in the ground truth (first row), most of the garment is blue (front side), with only small areas such as the collar and cuff being red (back side); but after processing (second and third rows), the garment of back and left view becomes mostly red, with only a small amount of blue remaining at the edges, and the front and back orientations are completely in chaos. Such errors are observed on both open and closed surface.

\noindent\textbf{Finding 2.} The face orientation becomes unreliable already at the 3D representation, i.e., the Flexible Dual Grid (FDG) stage, and the Shape-VAE reconstruction does not correct this issue.

\section{Approach}
Building on these findings, our goal is to inject correct face orientation into the pipeline while preserving generative priors of base model. To this end, we introduce \textit{FDG-D} (Flexible Dual Grid with Directed Edge, Sec.~\ref{sec:fdgd}), a minimal 3D representation that explicitly encodes normal direction via directed edges. To learn this representation, we propose \textit{ROS-FT} (Real Open Surface Finetuning, Sec.~\ref{sec:rosft}), a progressive post-training strategy. To prevent degradation of the original generative priors, we design a lightweight \textit{DE-Adapter} (Directed Edge Adapter, Sec.~\ref{sec:deadapter})---a parallel Feature Pyramid Network that leverages multi-scale global context to predict edge directions while keeping the base shape decoder frozen. Finally, to better benchmark performance, we construct \textit{Outfit3D} (Sec.~\ref{sec:outfit3d}), the first hybrid dataset comprising both open- and closed-surface 3D fashion assets.

\subsection{FDG-D}
\label{sec:fdgd}

Because the field-free 3D shape representation (FDG) in Trellis2~\cite{trellis2} discards explicit normals and gradients, face orientation inherently becomes undetermined. Our proposed \textit{FDG-D} resolves this by retaining normal direction from the original mesh, thereby unlocking the true capacity ceiling of Trellis2.

\textbf{FDG.} The original \textit{\mbox{Mesh}$\to$\mbox{FDG}} extraction entirely ignores normal information. During the \textit{\mbox{FDG}$\to$\mbox{mesh}} reconstruction, the winding order of each generated quad (i.e., its face orientation) is deterministically hardcoded based on the coordinate axis ($x$, $y$, or $z$) of its corresponding active edge. Because this winding is strictly bound to the spatial axis rather than the actual surface geometry, the algorithm cannot determine the true surface sidedness (front/back). Consequently, the extracted meshes suffer from arbitrary, chaotic face orientations, which is observed across open and closed surfaces.

\begin{figure}[t]
	\centering
	\includegraphics[width=\linewidth]{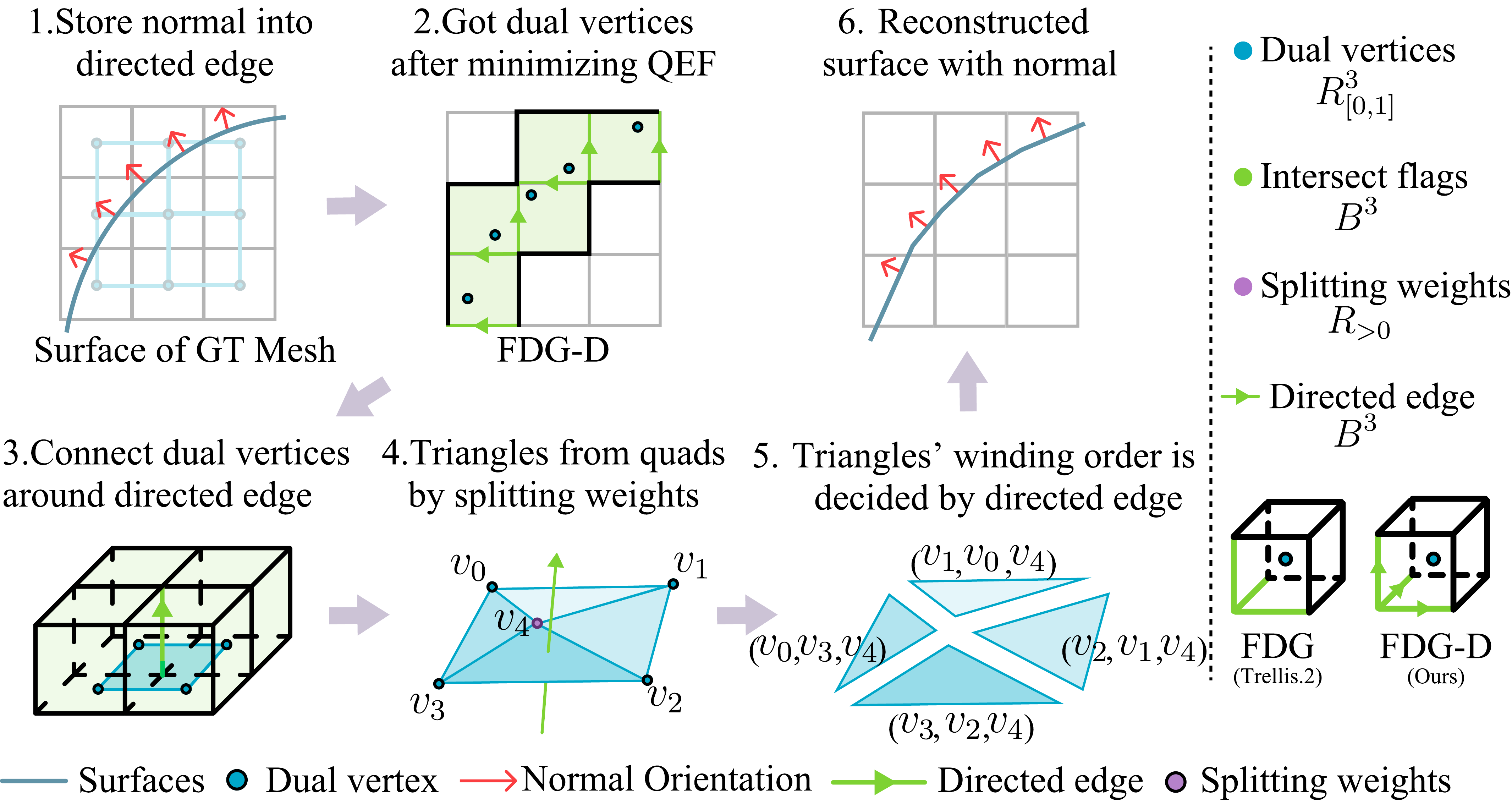}
	\caption{FDG-D (Flexible Dual Grid with Directed Edge) preserves the face orientation from GT Mesh in the directed edge, and uses it to orient the edge during the encoding (mesh $\to$ FDG-D), then determines the winding order by the edge orientation during the decoding (FDG-D $\to$ mesh).}
	\label{fig:FDG-D}
\end{figure}

\textbf{FDG-D.} To resolve the winding ambiguity, \textit{\mbox{Mesh}$\to$\mbox{FDG-D}} explicitly encodes the direction in which edge pierces the surface. For each active voxel, we store a 3-bit directional data corresponding to how its incident edges along the $X$, $Y$, and $Z$ axes intersect the mesh surface, compactly packed into a single \texttt{uint8} variable to minimize storage, inspired by intersect flags. To acquire these directions, we propose an \textit{Exact Edge-Ray Intersection} approach: each active dual-grid edge is treated as a ray to perform explicit ray-triangle intersection tests against the ground-truth mesh, yielding the exact direction of the edge as it penetrates the surface at the true intersection point. During the \textit{\mbox{FDG-D}$\to$\mbox{mesh}} reconstruction, the algorithm deterministically recovers the correct quad winding from these stored directions, as shown in the fourth step of Fig.~\ref{fig:FDG-D}. When generating a quad corresponding to an active edge, we evaluate whether its default geometric normal aligns with the expected crossing direction ($d_{\text{expected}} \in \{-1, 1\}$) retrieved from the FDG-D representation. If the default winding order contradicts the stored edge direction, we dynamically flip the quad's vertex indices (i.e., swapping the order from $[v_0, v_1, v_2, v_3]$ to $[v_0, v_3, v_2, v_1]$) prior to triangulation. This lightweight mechanism guarantees that the final reconstructed mesh strictly adheres to the right face orientation.

\begin{figure}[t]
	\centering
	\includegraphics[width=\linewidth]{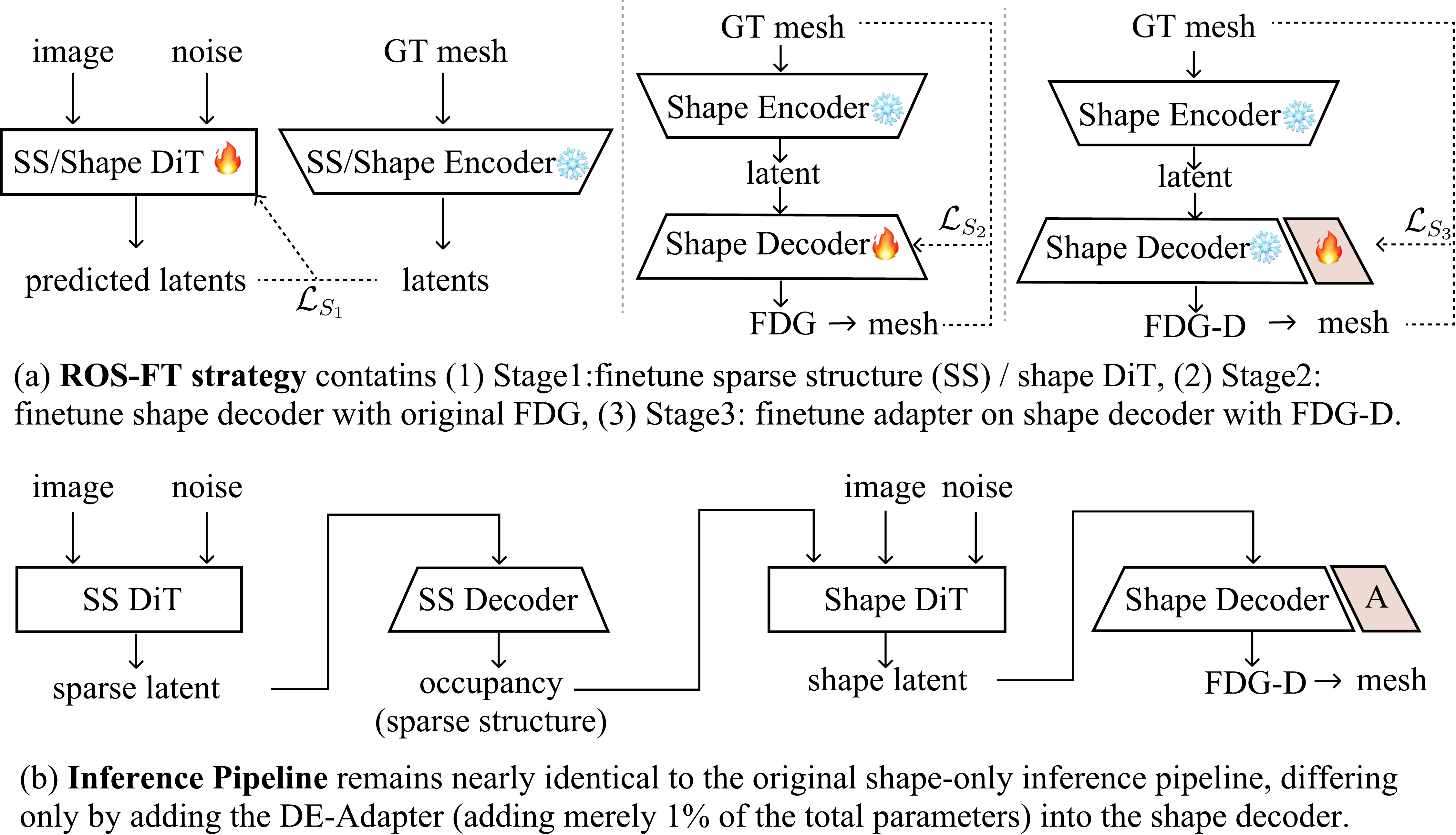}
	\caption{The (a) training and (b) inference pipeline of our method. A means Directed Edge Adapter (\textit{DE-Adapter}) on shape decoder.}
	\label{fig:ros_ft}
\end{figure}

\subsection{ROS-FT Strategy} 
\label{sec:rosft}
To unlock true open-surface generation capabilities---producing consistent face orientations---without sacrificing original performance on closed surfaces, we propose ROS-FT (Real Open Surface Finetuning). As shown in Fig.~\ref{fig:ros_ft}, ROS-FT is designed as a progressive, three-stage post-training pipeline: Stage 1 for DiT alignment, Stage 2 for FDG base reconstruction, and Stage 3 for DE-Adapter normal-aware tuning. Crucially, across all stages, we co-train on both open and closed surface data, using the latter as a regularization mechanism to preserve the model's foundational geometry priors.

\textbf{Stage 1: DiT Alignment.} 
The purpose of this stage is to align the latent generation capability with new data distribution. We finetune the rectified flow models (sparse structure DiT and shape DiT) using original forward process and optimize the neural network $\boldsymbol{v}_\theta$ via standard conditional flow matching (CFM) loss~\cite{lipman2023flow}:
\begin{equation} 
    \mathcal{L}_{S1}=\mathbb{E}_{t,\boldsymbol{x}_0,\boldsymbol{\epsilon}}\|\boldsymbol{v}_\theta(\boldsymbol{x}, t)-(\boldsymbol{\epsilon}-\boldsymbol{x}_0)\|^2_2. \label{eq:cfm} 
\end{equation} 

\textbf{Stage 2: FDG Base Reconstruction.} 
This stage aims to establish a robust foundation for decoding latents into explicit 3D geometry using the vanilla FDG representation. To strictly preserve the generative priors and avoid disrupting the latent space matched by the DiT pipeline, we keep the Shape VAE encoder completely frozen and solely fine-tune the decoder. The decoder is optimized using the original reconstruction objective $\mathcal{L}_{\text{s2}}$, which comprises structural, geometric, and rendering-based perceptual losses:

\begin{equation} 
  \mathcal{L}_{\text{s2}} = \lambda_{\text{v}}|\hat{\boldsymbol{v}}-\boldsymbol{v}|_2^2 + \lambda_{\delta}\operatorname{BCE}(\hat{\boldsymbol{\delta}},\boldsymbol{\delta}) + \lambda_{\boldsymbol{\rho}}\operatorname{BCE}(\hat{\boldsymbol{\rho}},\boldsymbol{\rho}) 
  + \lambda_{\text{KL}}\mathcal{L}_{\text{KL}} + \mathcal{L}_{\text{render}}. 
\end{equation} 

\textbf{Stage 3: DE-Adapter Tuning.} 
As the core of our strategy, Stage 3 addresses the critical orientation ambiguity inherent in the vanilla FDG. By the end of Stage 2, the Shape VAE has already been heavily constrained by 7 complex losses and predicts a 7-dimensional geometric output (including splitting weights for quad topology) based on a 6-dimensional input (dual vertices and intersected flags, as shown in Fig.~\ref{fig:FDG-D}). Introducing a new, explicit normal-orientation target directly into this base network severely destabilizes its pretrained topological capabilities---an optimization conflict we empirically verify in our ablation studies (see Tab.~\ref{tab:quant_ablation_fdg_d_vae_reconstruction} (c) and (d)).

To avoid this conflict and safely inject face orientation information without compromising the original Shape VAE's performance, Stage 3 introduces a lightweight Directed Edge Adapter (\textit{DE-Adapter}) branch alongside the strictly frozen base decoder. This adapter is tasked exclusively with predicting the crossing direction (the green vector in the 4th step of Fig.~\ref{fig:FDG-D}) logits $\hat{\boldsymbol{c}} \in \mathbb{B}^3$ for each spatial axis. To ensure the model focuses solely on valid geometry, the crossing direction loss is computed exclusively on the active voxels where the ground-truth intersection flag $\boldsymbol{\delta}$ is True. The supervision is formulated as a Binary Cross-Entropy with Logits loss against the ground-truth crossing direction $\boldsymbol{c} \in \{0, 1\}^3$:

\begin{equation}
    \mathcal{L}_{\text{s3}} =\lambda_{\text{s2}} \mathcal{L}_{\text{s2}} + \lambda_{\text{cd}} \cdot \operatorname{BCE}(\hat{\boldsymbol{c}}_{[\boldsymbol{\delta} = 1]}, \boldsymbol{c}_{[\boldsymbol{\delta} = 1]})
    \label{eq:stage3_loss}
\end{equation}
where $\lambda_{\text{s2}}$ and $\lambda_{\text{cd}}$ are weights of stage 2 loss and crossing direction loss, respectively. By restricting the trainable parameters to the adapter branch, Stage 3 effectively equips the decoder with precise face orientation awareness at a minimal computational cost.

\begin{figure*}[!t]
	\centering
	\includegraphics[width=\linewidth]{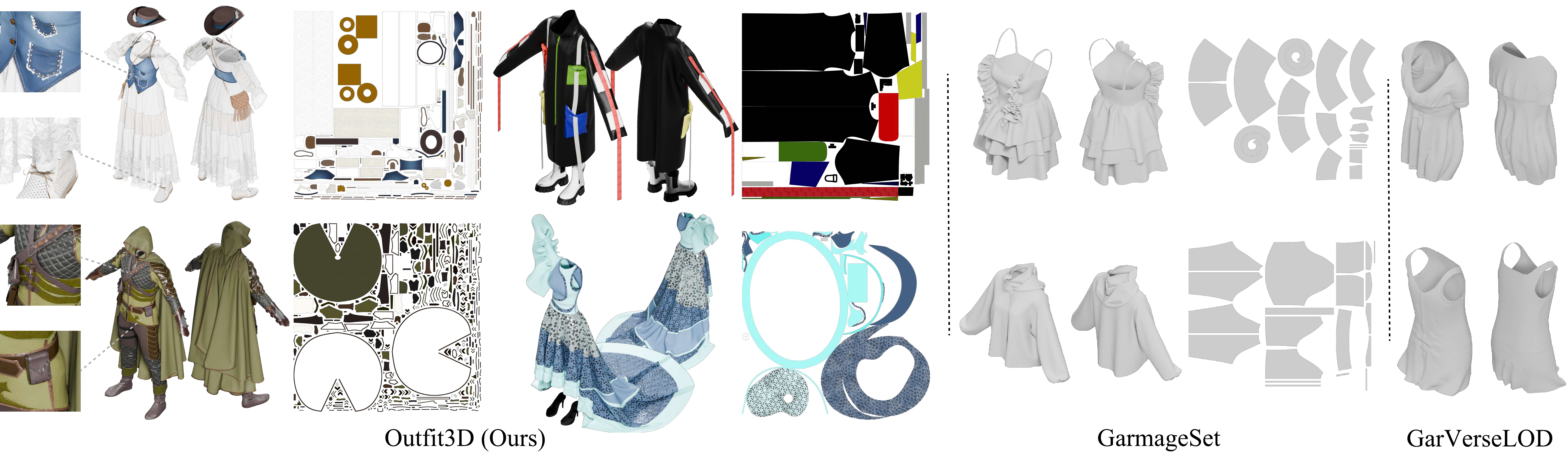}
    \vspace{-18pt}
	\caption{Examples of Outfit3D, GarmageSet, and GarVerseLOD.}
	\label{fig:4Data-compare}
    \vspace{-10pt}
\end{figure*}

\subsection{DE-Adapter}
\label{sec:deadapter}

To preserve the priors acquired during pre-training, our DE-Adapter (Directed Edge Adapter) is designed as a parallel Feature Pyramid Network (FPN) alongside the frozen shape decoder. Inspired by successful adapters in image generation~\cite{ye2023ipadapter, fonts, controlnet, WordCon}, this architecture allows the model to predict crossing edge directions without interfering with the base network. Predicting the correct direction fundamentally requires global context to determine the true surface sidedness (front/back), particularly in complex self-occlusion regions or thin garments. Therefore, the adapter extracts and processes intermediate features at every resolution stage from the shape decoder, fusing them with its own hidden state before upsampling. Crucially, it reuses the spatial subdivision topology ($\operatorname{subdiv}$) predicted by the base branch, ensuring the spatial alignment. This progressive, multi-scale integration grants the orientation prediction a global receptive field, significantly boosting the raw face orientation accuracy for complex open surfaces while maintaining parameter efficiency.

\subsection{Outfit3D Dataset}
\label{sec:outfit3d}

\begin{table}[t]
  \scriptsize
  \begin{tabular}{m{2.2cm}>{\centering\arraybackslash}m{2.4cm}>{\centering\arraybackslash}m{0.8cm}>{\centering\arraybackslash}m{1.2cm}}
  \toprule
  Methods & Surface Type & Texture & Multi-items \\
  \midrule
  GarmentCode {\color{gray}\scriptsize[ECCV 24]} & open surfac (garment only) & \tmarkno & \tmarkno \\
  GarVerseLOD {\color{gray}\scriptsize[TOG 24]} & open surface (garment only) & \tmarkno & \tmarkno \\
  GarmageSet {\color{gray}\scriptsize[TOG 25]} & open surface (garment only) & \tmarkno & \tmarkno \\
  Outfit3D (Ours)  & hybrid surface (outfits) & \tmarkok & \tmarkok \\
  \bottomrule
  \end{tabular}
  \caption{Differences with existing garment modeling datasets.}
  \label{tab:dataset_comparison}
\end{table}

Previous garment modeling datasets~\cite{korosteleva2024garmentcodedata,luo2024garverselod, zou2023cloth4d} primarily focus on isolated, open-surface garments. As summarized in Tab.~\ref{tab:dataset_comparison}, they lack the multi-item complexity inherent to real-world outfits. To bridge this gap and push the field toward realistic, production-level fashion generation, we construct \textit{Outfit3D}. As shown in Fig.~\ref{fig:4Data-compare}, this dataset provides complex, multi-item outfits that combine both open- and closed-surface components, complete with high-resolution textures.

\textit{Data Collection and Processing Pipeline.} 
The raw 3D assets in our dataset are sourced from open-source design communities and commercially purchased models, encompassing both standalone cloth assets and fully clothed human models. To ensure high-quality, production-ready geometry, we developed a rigorous three-step data curation pipeline:
(1) \textit{Garment Standardization via CLO3D}: Using the professional apparel design software CLO3D, we first standardize the UV layouts and bake all Physically Based Rendering (PBR) texture channels. The garments are then exported as high-fidelity thin-shell (open surface) meshes.
(2) \textit{Artifact Removal}: We utilize Blender scripts to automatically detect and delete non-garment auxiliary meshes from the exported models, such as transparent collision-bounding boxes for shoes and transparent simulation-assist patches on the garments.
(3) \textit{Avatar Removal and Export}: Finally, we strip the underlying human body meshes while preserving essential closed surface accessories (e.g., shoes, belts, and bags), exporting the finalized hybrid surface meshes for training. Further details of dataset statistics are provided in the supplementary material.

\section{Experiments}

\subsection{Implementation Details}
\textit{Training Configurations.} 
All experiments are conducted on 2 $\times$ H200 (141\,GiB) GPUs. The ROS-FT strategy is executed sequentially. In Stage 1, we finetune the sparse structure DiT and the shape DiT following the original settings, utilizing a batch size of 24 with learning rates of \(1\times10^{-4}\) and \(2\times10^{-5}\), respectively, using the AdamW optimizer. In Stage 2, we freeze the shape encoder and finetune the shape decoder using a batch size of 48 and a learning rate of \(1\times10^{-5}\). Finally, in Stage 3, we freeze the entire base VAE and train the proposed DE-Adapter with a learning rate of \(1\times10^{-3}\).

\textit{Dataset Setup.} 
To construct a comprehensive training and evaluation benchmark, we compile a mixed dataset containing 10,489 assets. This collection comprises 3,724 single-item open surface garments from GarmageSet~\cite{li2025garmagenet}, 3,266 outfits from our proposed Outfit3D dataset, and 3,499 closed surface objects from ObjaverseXL~\cite{deitke2023objaversexl}. The inclusion of the ObjaverseXL data acts as a regularization mechanism to preserve the pre-trained model's general 3D generation capabilities. From each of these three domains, we randomly reserve 100 assets (300 in total) for evaluation, leaving the remaining 10,189 assets for training.

\textit{Ablation and Inference Settings.} 
To accelerate experimental iterations under limited computing resources, our ablation studies regarding FDG-D representation (for Shape VAE reconstruction and shape generation, i.e, Tab.~\ref{tab:quant_ablation_fdg_d_vae_reconstruction} and Tab.~\ref{tab:quant_ablation_fdg_d_shape_generation}) are conducted on a lightweight subset. We randomly sample 500 assets from GarmageSet~\cite{li2025garmagenet} for training and another 50 for evaluation. Furthermore, for quantitative evaluation in shape generation, we deliberately disable post-processing steps (such as remeshing) from the original Trellis2 inference pipeline, retaining only hole-filling. This ensures that the reported metrics strictly reflect the original normal quality of the generated mesh without external refinement.

\subsection{Evaluation Metrics} 
\textit{Geometry Accuracy.} 
To comprehensively evaluate geometric fidelity, we employ three standard metrics following established protocols~\cite{linpartcrafter, yang2025omnipart}. First, Chamfer Distance (CD) is used to measure the bidirectional point-to-point surface deviation between the generated meshes and the ground truth by uniformly sampling $10^5$ points on the surfaces. Second, we report the F-Score at a specified distance threshold ($\tau=0.03$ m) to evaluate the percentage of correctly reconstructed surfaces, capturing both precision and recall. Finally, Voxel IoU (Intersection over Union) is computed on a rasterized $64^3$ grid to assess the global structural preservation. 

\textit{Normal Quality.} 
Face orientation is a critical focus of our work. Following NeuralGF~\cite{li2023neuralgf}, we report two Root Mean Square Error (RMSE) metrics between the ground-truth normals $\hat{\boldsymbol{n}}_i$ and the estimated normals $\boldsymbol{n}_i$: the unoriented error \(RMSE_U=\sqrt{\frac{1}{I}\sum_{i=1}^{I}\left(\arccos\left(\left|\hat{\boldsymbol{n}}_i \odot \boldsymbol{n}_i\right|\right)\right)^2}\) and the oriented error \(RMSE_O = \\ \sqrt{\frac{1}{I}\sum_{i=1}^{I}\left(\arccos\left(\hat{\boldsymbol{n}}_i \odot \boldsymbol{n}_i\right)\right)^2}\). Here, $RMSE_U\!\in\![0^\circ,90^\circ]$ and $RMSE_O\!\in\![0^\circ,180^\circ]$; $I$ is the number of evaluated normals ($10^4$ points sampled on the prediction mesh queried against the nearest ground-truth triangles), $\odot$ denotes the inner product, and $|\cdot|$ denotes the absolute value. Crucially, we also report the orientation error rate $\tau_o(\%)$, computed as $100-\textit{Correct orientation}(\%)$. A lower $\tau_o$ indicates better inside-outside differentiation, allowing a global flip of predicted normals if it yields a lower error to account for global ambiguity.

\textit{Topology Quality.} 
We follow DoubleCoverUDF~\cite{hou2023robust} to assess topological robustness by percentage of non-manifold vertices $\tau_v(\%)$, the number of boundary edges $N_b$, the number of open edge circles (boundary loops) $N_c$, and the genus $N_g$. Non-manifold vertices occur when local neighborhoods intersect at a single point (forming multiple topological disks), which we conservatively identify via edges shared by more than two triangles. These defects are highly undesirable as they destabilize downstream simulations, and forcefully removing them often introduces unwanted holes. $N_b$, $N_c$, and $N_g$ capture the global topological features. The genus $N_g$ is derived directly from the Euler characteristic of generated mesh.

\subsection{Comparison}
We comprehensively evaluate our approach across three domains: pure open, pure closed, and hybrid surfaces. We first compare against garment-specific baselines (ChatGarment~\cite{bian2025chatgarment}, Design2GarmentCode~\cite{zhou2025design2garmentcode}, and SewingLDM~\cite{liu2025multimodal}), which synthesize meshes by predicting 2D patterns and stitching them via physical simulation. Because these methods cannot generate pure closed surfaces, we test them only on GarmageSet and Outfit3D datasets. Furthermore, while their pattern-stitching paradigm inherently prevents non-manifold vertices and genus artifacts (making direct topology comparisons inapplicable), it introduces severe simulation bottlenecks: on our evaluation set, SewingLDM only simulated 28 cases successfully, and Design2GarmentCode failed on 2 cases. In contrast, our AnySurf pipeline directly generates any surface with a 100\% success rate.

\subsubsection{Quantitative Comparison}
\textit{Performance on open surfaces.} We first evaluate the models on the open surface domain (GarmageSet) (Tab.~\ref{tab:quant_compare_open_surface}). While garment-specific models (ChatGarment, Design2GarmentCode, SewingLDM) are natively designed for this domain, their pattern-stitching paradigm struggles to accurately reconstruct complex 3D shapes, resulting in lower geometric fidelity (e.g., lower F-Score and Voxel IoU) and poor face orientation ($\tau_o \approx 84-85\%$). Compared to general 3D generators, the original Trellis2 baseline exhibits catastrophic normal flipping ($\tau_o = 52.29\%$) and severe topological defects (over 1900 open edge circles). Directly fine-tuning Trellis2 significantly improves geometry but fundamentally fails to resolve the normal ambiguity ($\tau_o = 51.71\%$). In contrast, our full AnySurf pipeline rectifies this deficiency, achieving state-of-the-art normal quality ($\tau_o = 90.39\%$) while preserving the highest geometric accuracy ($IoU=0.7674$) and topological robustness.

\textit{Performance on hybrid surfaces.} We further extend our evaluation to the more challenging hybrid surface from our Outfit3D dataset. Outfit3D presents significantly higher topological complexity than GarmageSet: intricate structures like lace and complex seams (Fig.~\ref{fig:4Data-compare}) surge its average GT open boundary circles ($N_c$) to 2512.7, compared to GarmageSet's 16.9. As shown in Tab.~\ref{tab:quant_compare_hybrid_data}, the presence of mixed open and closed topologies exposes the limitations of existing methods. Garment-specific baselines suffer a severe performance drop when forced to handle hybrid structures (e.g., outfits containing solid shoes or thick belts). Their geometric fidelity degrades significantly (IoU drops to $\approx 0.12-0.15$, and CD surges to $\approx 0.13-0.17$, an order of magnitude higher than ours), confirming that 2D sewing-pattern paradigms cannot generalize to real-world outfits with closed volumetric components. On the other hand, the general-purpose Trellis2 baseline maintains decent geometric accuracy but completely fails to differentiate the front side from the back side ($\tau_o = 52.82\%$). While fine-tuning Trellis2 on hybrid data improves geometric alignment ($IoU = 0.7220$), it remains blind to orientation ($\tau_o = 52.31\%$). Ours successfully bridges this gap and achieves the best normal quality ($\tau_o = 86.21\%$) and the lowest oriented error ($\mathrm{RMSE}_O = 57.035^\circ$), while simultaneously delivering the highest geometric accuracy ($CD=0.0164$) and robust topology ($\tau_v=1.3607\%$).

\begin{table*}[t]
  \centering
  \setlength{\tabcolsep}{4pt}
       \begin{tabularx}{\textwidth}{l
          >{\hsize=0.75\hsize\centering\arraybackslash}X
          >{\hsize=0.75\hsize\centering\arraybackslash}X
          >{\hsize=0.75\hsize\centering\arraybackslash}X
          >{\hsize=0.75\hsize\centering\arraybackslash}X
          >{\hsize=0.75\hsize\centering\arraybackslash}X
          >{\hsize=0.75\hsize\centering\arraybackslash}X
          >{\hsize=0.75\hsize\centering\arraybackslash}X
          >{\hsize=0.75\hsize\centering\arraybackslash}X
          >{\hsize=0.75\hsize\centering\arraybackslash}X}
          \toprule
          \multirow{2}{*}{Methods} & \multicolumn{3}{c}{Geometry Accuracy} & \multicolumn{3}{c}{Normal Quality} & \multicolumn{3}{c}{Topology Quality} \\
          \cmidrule(lr){2-4} \cmidrule(lr){5-7} \cmidrule(lr){8-10}
          ~ & CD${\downarrow}$ & F-Score${\uparrow}$ & IoU${\uparrow}$ & $RMSE_U{\downarrow}$ & $RMSE_O{\downarrow}$ & $\tau_o(\%){\uparrow}$  & $N_c{\downarrow}$ & $N_g{\downarrow}$ & $\tau_v(\%){\downarrow}$ \\
          \midrule
          Ground Truth & 0.003557 & 1.0000 & 1.0000 & 6.021 & 7.953 & 99.80 & 16.9 & 29.4 & 0.0036 \\
          ChatGarment & 0.058705 & 0.9620 & 0.3147 & 42.351 & 59.520 & 85.24 & - & - & - \\
          Design2GarmentCode & 0.049524 & 0.9703 & 0.3703 & 43.587 & 61.039 & 84.44 & - & - & - \\
          SewingLDM & 0.072085 & 0.9483 & 0.2496 & 46.007 & 60.607 & 84.67 & - & - & - \\
          Trellis2 & 0.024585 & 0.9287 & 0.5842 & 33.868 & 109.798 & 52.29 & 1909.1 & 1325.6 & 1.9782 \\
          Trellis2 (Finetuned) & \textbf{0.012772} & \textbf{0.9904} & \textbf{0.7674} & 23.533 & 115.530 & 51.71 & 122.9 & 23.4 & 0.1801 \\
          Ours & 0.012774 & \textbf{0.9904} & \textbf{0.7674} & \textbf{21.679} & \textbf{48.112} & \textbf{90.39} & \textbf{121.2} & \textbf{21.8} & \textbf{0.1771} \\
          \bottomrule 
      \end{tabularx}
  \caption{Quantitative comparison of our method with state-of-the-art models on pure open surface (GarmageSet) generation task.}
  \label{tab:quant_compare_open_surface}
\end{table*}

\begin{table*}[t]
  \centering
  \setlength{\tabcolsep}{4pt}
       \begin{tabularx}{\textwidth}{l
          >{\hsize=0.75\hsize\centering\arraybackslash}X
          >{\hsize=0.75\hsize\centering\arraybackslash}X
          >{\hsize=0.75\hsize\centering\arraybackslash}X
          >{\hsize=0.75\hsize\centering\arraybackslash}X
          >{\hsize=0.75\hsize\centering\arraybackslash}X
          >{\hsize=0.75\hsize\centering\arraybackslash}X
          >{\hsize=0.75\hsize\centering\arraybackslash}X
          >{\hsize=0.75\hsize\centering\arraybackslash}X
          >{\hsize=0.75\hsize\centering\arraybackslash}X}
          \toprule
          \multirow{2}{*}{Methods} & \multicolumn{3}{c}{Geometry Accuracy} & \multicolumn{3}{c}{Normal Quality} & \multicolumn{3}{c}{Topology Quality} \\
          \cmidrule(lr){2-4} \cmidrule(lr){5-7} \cmidrule(lr){8-10}
          ~ & CD${\downarrow}$ & F-Score${\uparrow}$ & IoU${\uparrow}$ & $RMSE_U{\downarrow}$ & $RMSE_O{\downarrow}$ & $\tau_o(\%){\uparrow}$  & $N_c{\downarrow}$ & $N_g{\downarrow}$ & $\tau_v(\%){\downarrow}$ \\
          \midrule
          Ground Truth & 0.002929 & 1.0000 & 1.0000 & 13.371 & 24.576 & 97.53 & 2512.7 & 9.3 & 2.4926\\
          ChatGarment & 0.136093 & 0.7725 & 0.1436 & 46.220 & 63.220 & 82.70 & - & - & - \\
          Design2GarmentCode & 0.142790 & 0.7436 & 0.1509 & 46.412 & 64.635 & 81.31 &  - & - & - \\
          SewingLDM & 0.176955 & 0.6040 & 0.1244 & 45.945 & 63.385 & 84.13 & - & - & - \\
          Trellis2 & 0.023919 & 0.9354 & 0.6194 & 36.700 & 108.241 & 52.82 & 2132.6 & 2298.0 & 2.8546\\
          Trellis2 (Finetuned) & 0.017087 & \textbf{0.9672} & 0.7220 & 32.572 & 110.986 & 52.31 & 781.8 & 110.7 & 1.3609 \\
          Ours & \textbf{0.016433} & \textbf{0.9672} & \textbf{0.7309} & \textbf{29.645} & \textbf{57.035} & \textbf{86.21} & \textbf{781.7} & \textbf{109.0} & \textbf{1.3607}\\
          \bottomrule 
      \end{tabularx}
  \caption{Quantitative comparison of our method with state-of-the-art models on hybrid surface data (Outfit3D) generation task.}
  \label{tab:quant_compare_hybrid_data}
\end{table*}

\begin{figure*}[t]
  \centering
  \includegraphics[width=0.95\linewidth]{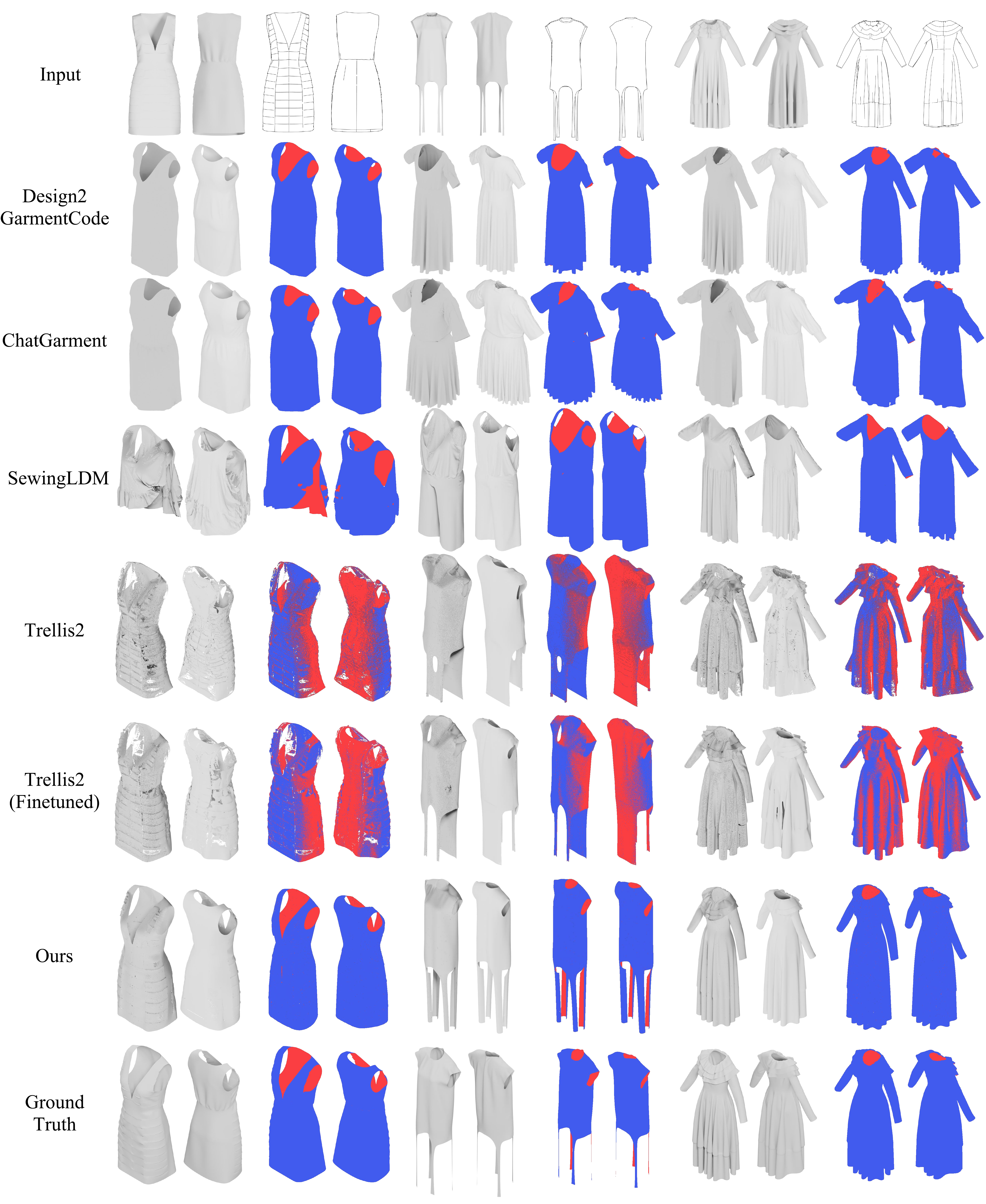}
  \vspace{-12pt}
  \caption{Qualitative comparison on pure open surface (GarmageSet) generation task.}
  \label{fig:5compare-open}
  \vspace{-8pt}
\end{figure*}

\begin{figure*}[t]
  \centering
  \includegraphics[width=0.85\linewidth]{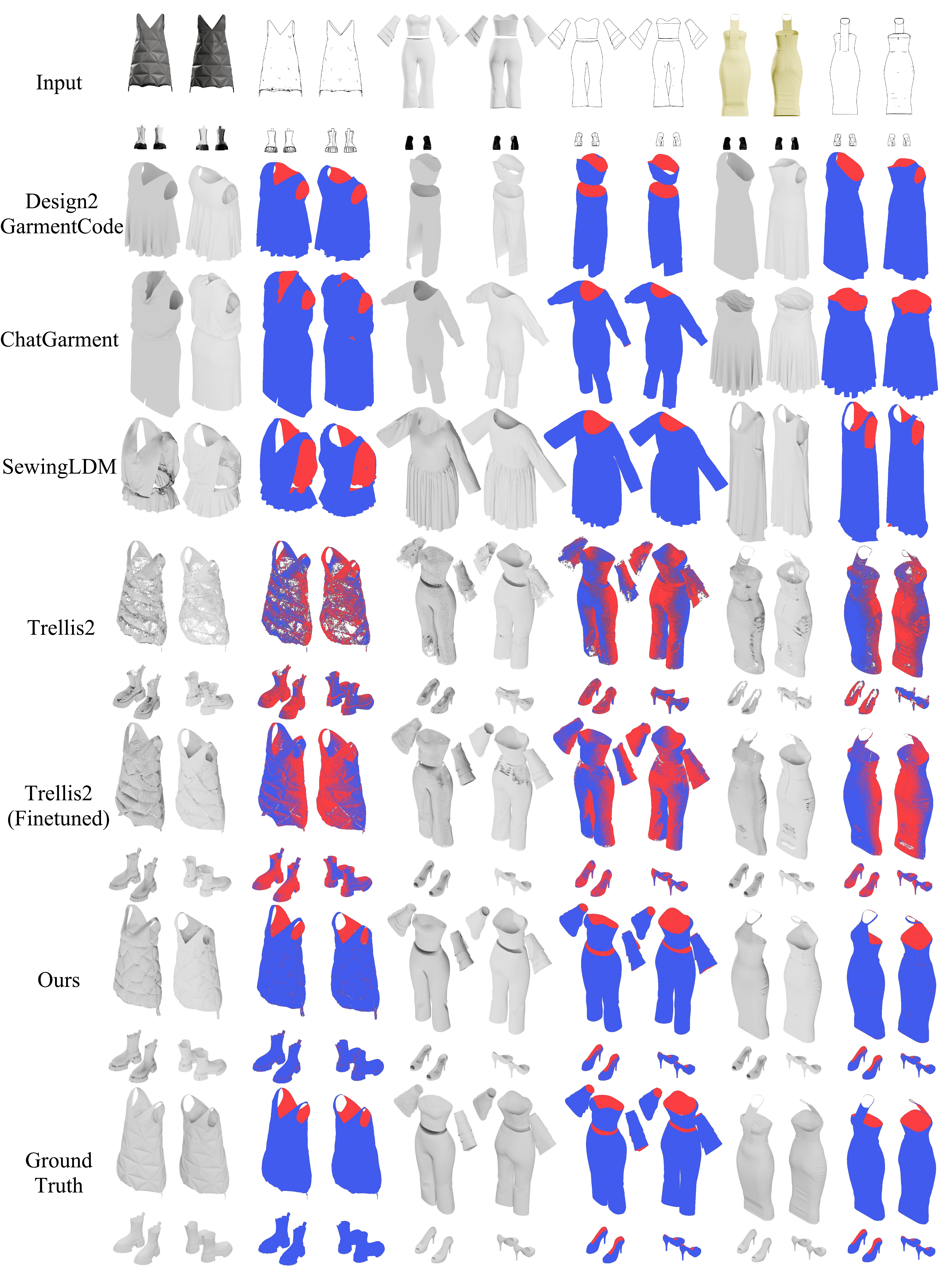}
  \vspace{-12pt}
  \caption{Qualitative comparison on hybrid data (Outfit3D) generation task.}
  \label{fig:5compare-hybrid}
  \vspace{-8pt}
\end{figure*}

\subsubsection{Qualitative Comparison}

SewingLDM does not support rendered image conditioning; we therefore use its sketch-based input for that model and rendered images for all other methods under otherwise identical evaluation settings. Fig.~\ref{fig:5compare-open} shows qualitative results on \textit{open surfaces} (GarmageSet). Garment-specific baselines (ChatGarment, Design2GarmentCode, and SewingLDM) mostly preserve coherent inside--outside shading (blue exteriors and red interiors), but their sewing-pattern pipelines cap geometric complexity and yield overly simplified shapes. The general-purpose Trellis2 baseline instead produces fragmented meshes and severe normal noise, with red and blue interleaved on outward-facing regions. Fine-tuning Trellis2 improves geometric fidelity yet leaves orientation largely unresolved. In contrast, our method recovers intricate wrinkles and folds and attains clean, globally consistent face orientations that closely align with the ground truth.

\begin{figure}[t]
  \centering
  \includegraphics[width=\linewidth]{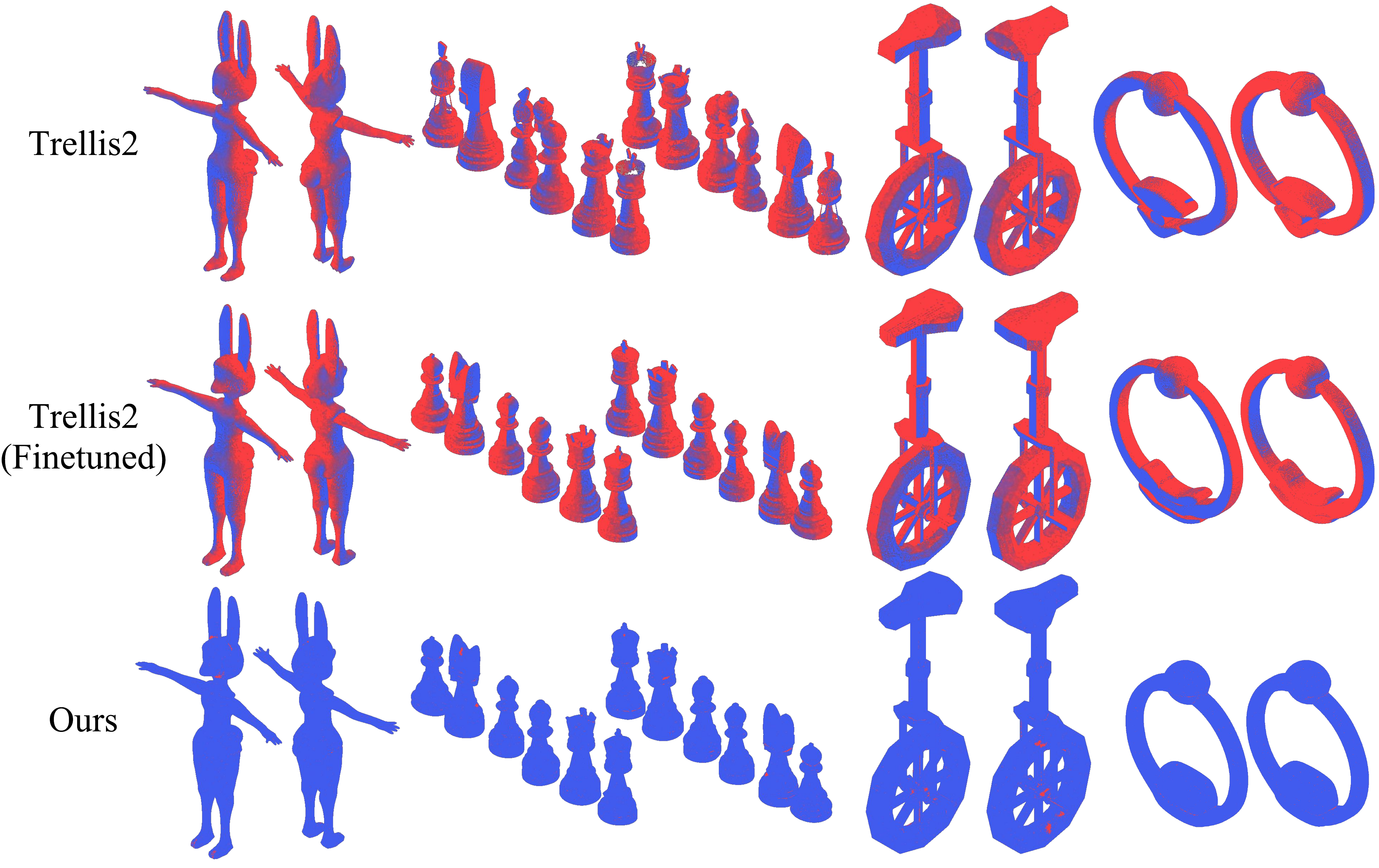}
  \vspace{-12pt}
  \caption{Qualitative comparison on pure closed surface generation task.}
  \label{fig:5compare-closed}
  \vspace{-5pt}
\end{figure}

\textit{Performance on hybrid surfaces.} We further compare visual quality on the more challenging Outfit3D dataset (Fig.~\ref{fig:5compare-hybrid}), which mixes open (e.g., dresses) and closed (e.g., shoes) topologies. Garment-specific baselines struggle significantly here: they completely fail to generate closed components like shoes, and their outputs for the main garments are overly simplistic, lacking the fidelity to align with the input images. While the general-purpose Trellis2 can generate all components, it produces severely corrupted geometries and chaotic normals, with large patches of red incorrectly appearing on the exterior of both garments and shoes. Fine-tuning Trellis2 improves structural coherence, but the normal flipping issue persists, particularly on complex regions and closed structures. In contrast, our method robustly handles hybrid topologies, generating highly detailed geometries for both open garments and closed accessories while consistently maintaining correct face orientations across all components, achieving results align with the ground truth.

\textit{Performance on closed surfaces.} Fig.~\ref{fig:5compare-closed} shows closed surface examples (e.g., characters, vehicles); garment-specific baselines do not apply, so we compare only to Trellis2 and its finetuned variant. Both still exhibit red--blue interleaving artifacts remains unreliable even on watertight meshes. AnySurf outperforms them with coherent orientation (uniform blue) without degrading shape fidelity.

\subsection{Ablation Study}

\subsubsection{Ablation of FDG-D}

\begin{figure}[t]
	\centering
	\includegraphics[width=\linewidth]{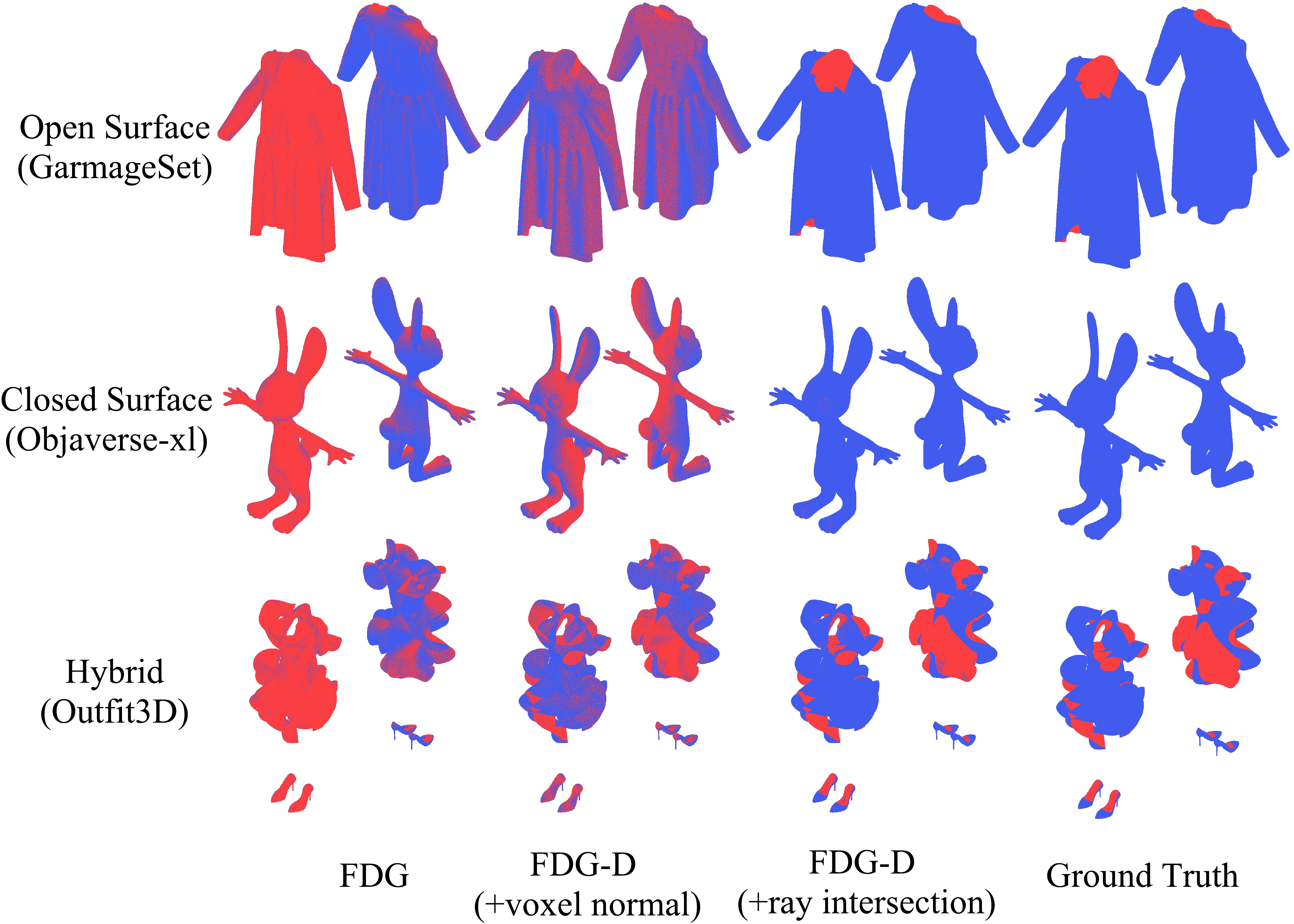}
	\caption{Ablation study of FDG-D on 3D representation reconstruction task. FDG-D with face normal guidance can significantly improve the normal quality to the ground truth level even in very challenging case (the last row) and ignoring if open surface data or closed surface data.}
	\label{fig:Ablation-1}
\end{figure}

On the \textbf{3D representation reconstruction} task (mesh $\to$ FDG-D $\to$ mesh), we evaluate the intrinsic fidelity of the representation itself without any neural network intervention. We use 300 instances in total, with 100 drawn from each of GarmageSet~\cite{li2025garmagenet}, Objaverse-XL~\cite{deitke2023objaverse}, and our Outfit3D dataset. The quantitative results are reported in Tab.~\ref{tab:quant_ablation_fdg_d_3d_rep}. The baseline FDG suffers from severe normal flipping ($\tau_o \approx 51.62\%$, near random guessing) due to the lack of explicit orientation encoding, which visually manifests as chaotic red/blue surface patches (see Fig.~\ref{fig:Ablation-1}). To address this, we ablate two variants of normal assignment for our FDG-D representation: (a) \textit{Voxel Normal Guidance}, which computes and stores the average normal of all surfaces intersecting a voxel; and (b) \textit{Ray Intersection}, which explicitly stores the exact crossing direction of the active edges based on explicit ray-triangle intersection tests against the ground-truth mesh. As shown in Tab.~\ref{tab:quant_ablation_fdg_d_3d_rep} and qualitatively corroborated in Fig.~\ref{fig:Ablation-1}, voxel-level averaging yields only marginal improvements. In contrast, storing the exact edge-ray intersection directions explicitly binds orientation to the topology, significantly boosting the correct orientation rate $\tau_o$ to $95.04\%$ and drastically reducing the oriented error ($RMSE_O$) from $121.7$ to $31.7$, successfully restoring uniform, ground-truth-level normals across open, closed, and hybrid surfaces, all without compromising geometry or topology accuracy.

\begin{figure}[t]
	\centering
	\includegraphics[width=\linewidth]{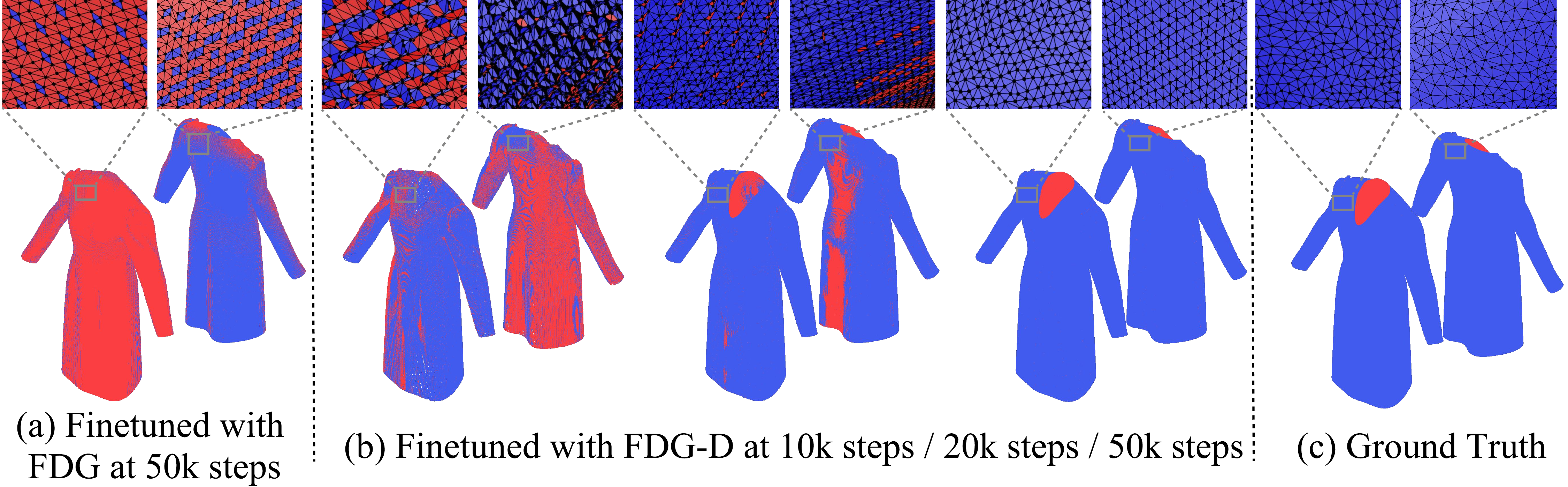}
	\caption{Ablation study of FDG-D on shape VAE reconstruction task. Shape decoder finetuned with FDG-D can gradually improve the normal quality compared to the baseline FDG still limit in normal quality.}
	\label{fig:Ablation-2}
\end{figure}

On the \textbf{shape VAE reconstruction} task (mesh $\to$ FDG-D $\to$ Shape VAE $\to$ FDG-D $\to$ mesh), to accelerate experimental iterations under limited computing resources, we only use 500 instances from GarmageSet~\cite{li2025garmagenet}. And we train the Shape VAE decoder while freezing the encoder for keeping the latent space as same as DiT. The quantitative results are reported in Tab.~\ref{tab:quant_ablation_fdg_d_vae_reconstruction}. To systematically validate our DE-Adapter design, we perform extensive ablations on the network architecture and training strategies. 
First, we investigate the source of edge direction during the decoding phase (FDG-D $\to$ mesh). As shown in variants (c) and (d), using ground-truth orientations (d) instead of predicted ones (c) yields similar performance. However, to maintain consistency with the topological splitting weights (which rely on ground truth during decoding), we adopt GT orientations for subsequent ablations.
Second, we demonstrate that directly fine-tuning the base decoder on the FDG-D representation (variants c, d, g) successfully improves normal quality but inevitably degrades the topology quality (e.g., $N_c$ and $N_g$ significantly increase compared to the baseline). 
Third, we evaluate the addition of a simple linear head (variants e, f). While this two-stage approach preserves the pretrained topology, the shallow head lacks the capacity to learn complex edge directions even with maximum loss weights, resulting in a low $\tau_o$ of $58.93\%$.
Finally, our introduced DE-Adapter (variant h, with $\mathcal{L}_{\text{s2}}=0$) resolves this dilemma. The model successfully learns the correct face orientation ($\tau_o=90.67\%$ at depth 32) while strictly guaranteeing that the topological quality remains completely unaffected compared to the finetuned baseline. Notably, increasing the complexity of the adapter (scaling the depth from 8 to 32) enhances normal quality without compromising either geometric accuracy or topology quality. Furthermore, Fig.~\ref{fig:Ablation-2} reveals that while the VAE fails to capture normal directionality using the original FDG, our FDG-D representation effectively enables the model to acquire this capability.

\begin{table*}[t]
  \centering
  \setlength{\tabcolsep}{4pt}
       \begin{tabularx}{\textwidth}{l
          >{\hsize=0.75\hsize\centering\arraybackslash}X
          >{\hsize=0.75\hsize\centering\arraybackslash}X
          >{\hsize=0.75\hsize\centering\arraybackslash}X
          >{\hsize=0.75\hsize\centering\arraybackslash}X
          >{\hsize=0.75\hsize\centering\arraybackslash}X
          >{\hsize=0.75\hsize\centering\arraybackslash}X
          >{\hsize=0.75\hsize\centering\arraybackslash}X
          >{\hsize=0.75\hsize\centering\arraybackslash}X
          >{\hsize=0.75\hsize\centering\arraybackslash}X}
          \toprule
          \multirow{2}{*}{Methods} & \multicolumn{3}{c}{Geometry Accuracy} & \multicolumn{3}{c}{Normal Quality} & \multicolumn{3}{c}{Topology Quality} \\
          \cmidrule(lr){2-4} \cmidrule(lr){5-7} \cmidrule(lr){8-10}
          ~ & CD${\downarrow}$ & F-Score${\uparrow}$ & IoU${\uparrow}$ & $RMSE_U{\downarrow}$ & $RMSE_O{\downarrow}$ & $\tau_o(\%){\uparrow}$  & $N_c{\downarrow}$ & $N_g{\downarrow}$ & $\tau_v(\%){\downarrow}$ \\
          \midrule 
          Ground Truth & 0.003588 & 1.000 & 1.000 & 12.574 & 22.901 & 96.88 & 854.3 & 13.2 & 1.4328\\
          FDG (Baseline) & 0.003614  & 0.999  & 0.964 & 14.221 & 121.734 & 51.62 & 988.6 & 192.6 & 2.8959\\
          FDG-D (a) & 0.003613  & 0.999  & 0.964  & 14.215 & 106.353 & 62.19 & 988.7 & 192.6 & 2.8958\\
          FDG-D (b) & 0.003614  & 0.999  & 0.964  & 14.192 & 31.775 & 95.04 & 988.6 & 192.6 & 2.8958\\
          \bottomrule 
      \end{tabularx}
  \caption{Ablation study of FDG-D on 3D representation reconstruction with different settings: (a) use voxel normal guidance and (b) use edge-ray intersection.}
\label{tab:quant_ablation_fdg_d_3d_rep}
\end{table*}

\begin{table*}[t]
  \centering
  \setlength{\tabcolsep}{4pt}
       \begin{tabularx}{\textwidth}{l
          >{\hsize=0.75\hsize\centering\arraybackslash}X
          >{\hsize=0.75\hsize\centering\arraybackslash}X
          >{\hsize=0.75\hsize\centering\arraybackslash}X
          >{\hsize=0.75\hsize\centering\arraybackslash}X
          >{\hsize=0.75\hsize\centering\arraybackslash}X
          >{\hsize=0.75\hsize\centering\arraybackslash}X
          >{\hsize=0.75\hsize\centering\arraybackslash}X
          >{\hsize=0.75\hsize\centering\arraybackslash}X
          >{\hsize=0.75\hsize\centering\arraybackslash}X}
          \toprule
          \multirow{2}{*}{Methods} & \multicolumn{3}{c}{Geometry Accuracy} & \multicolumn{3}{c}{Normal Quality} & \multicolumn{3}{c}{Topology Quality} \\
          \cmidrule(lr){2-4} \cmidrule(lr){5-7} \cmidrule(lr){8-10}
          ~ & CD${\downarrow}$ & F-Score${\uparrow}$ & IoU${\uparrow}$ & $RMSE_U{\downarrow}$ & $RMSE_O{\downarrow}$ & $\tau_o(\%){\uparrow}$  & $N_c{\downarrow}$ & $N_g{\downarrow}$ & $\tau_v(\%){\downarrow}$ \\
          \midrule 
          Ground Truth & 0.003584 & 1.000 & 1.000 & 6.017 & 7.980 & 99.77 & 17.8 & 32.4 & 0.0038\\
          Pretrained decoder + FDG & 0.003588 & 1.000 & 0.9698 & 6.543 & 122.286 & 52.20 & 707.9 & 121.9 & 0.6805\\
          Finetuned decoder + FDG (Baseline) & 0.003588 & 1.000 & 0.9695 & 6.373 & 122.230 & 52.22 & 718.8 & 190.5 & 0.6128\\
          Finetuned decoder + FDG-D (c)-0.1 & 0.003591 & 1.000 & 0.9698 & 7.274 & 20.558 & 97.75 & 1050.7 & 240.1 & 1.0397\\
          Finetuned decoder + FDG-D (d)-0.1 & 0.003592 & 1.000 & 0.9698 & 7.209 & 20.154 & 97.88 & 1052.6 & 238.8 & 1.0409\\
          Finetuned decoder + FDG-D (d)-0.05 & 0.003594 & 1.000 & 0.9697 & 7.215 & 22.302 & 97.46 & 1035.3 & 235.6 & 1.0387\\
          Finetuned decoder + FDG-D (d)-0.01 & 0.003594 & 1.000 & 0.9696 & 7.165 & 35.962 & 94.90 & 999.7 & 242.0 & 1.0066\\
          Finetuned decoder + FDG-D (e)-0.01 & 0.003588 & 1.000 & 0.9695 & 6.381 & 113.380 & 58.83 & 718.8 & 190.5 & 0.6128\\
          Finetuned decoder + FDG-D (f) & 0.003587 & 1.000 & 0.9695 & 6.400 & 113.231 & 58.93 & 718.8 & 190.5 & 0.6128\\
          Finetuned decoder + FDG-D (g)-0.01 & 0.003592 & 1.000 & 0.9686 & 9.124 & 30.897 & 95.94 & 812.9 & 220.6 & 0.6833\\
          Finetuned decoder + FDG-D (h)-8 & 0.003587 & 1.000 & 0.9695 & 6.360 & 63.713 & 86.22 & 718.7 & 190.3 & 0.6128\\
          Finetuned decoder + FDG-D (h)-32 & 0.003585 & 1.000 & 0.9695 & 6.395 & 51.016 & 90.67 & 718.7 & 190.3 & 0.6128\\
          \bottomrule 
      \end{tabularx}
  \caption{Ablation of FDG-D on Shape VAE reconstruction. \textbf{Edge-orientation source during decoding:} (c)~predicted orientations from the shape decoder; (d)~ground-truth orientations from the mesh. \textbf{Two-stage training for decoder (stage~2$\rightarrow$stage~3):} (e)/(f)~only the added edge-orientation head is updated in stage~3; (g)~both the new head and full decoder are updated in stage~2\&3; (h)~the added adapter is updated in stage~3. \textbf{Loss:} (c), (d), (e), and (g) use the original loss $\mathcal{L}_{\text{s2}}$ plus $\mathcal{L}_{\mathrm{cd}}$; (f)/(h) uses $\mathcal{L}_{\mathrm{cd}}$ only. The weight $\lambda_{\mathrm{cd}}$ of $\mathcal{L}_{\mathrm{cd}}$: (c)~$\lambda_{\mathrm{cd}}{=}0.1$; (d)~$\lambda_{\mathrm{cd}}{\in}\{0.1,0.05,0.01\}$; (e)--(g)~$\lambda_{\mathrm{cd}}{=}0.01$. (h)-8/32: depth of adapter(v1)=8/32.}
  \label{tab:quant_ablation_fdg_d_vae_reconstruction}
\end{table*}

\begin{table*}[t]
  \centering
  \setlength{\tabcolsep}{4pt}
       \begin{tabularx}{\textwidth}{l
          >{\hsize=0.75\hsize\centering\arraybackslash}X
          >{\hsize=0.75\hsize\centering\arraybackslash}X
          >{\hsize=0.75\hsize\centering\arraybackslash}X
          >{\hsize=0.75\hsize\centering\arraybackslash}X
          >{\hsize=0.75\hsize\centering\arraybackslash}X
          >{\hsize=0.75\hsize\centering\arraybackslash}X
          >{\hsize=0.75\hsize\centering\arraybackslash}X
          >{\hsize=0.75\hsize\centering\arraybackslash}X
          >{\hsize=0.75\hsize\centering\arraybackslash}X}
          \toprule
          \multirow{2}{*}{Methods} & \multicolumn{3}{c}{Geometry Accuracy} & \multicolumn{3}{c}{Normal Quality} & \multicolumn{3}{c}{Topology Quality} \\
          \cmidrule(lr){2-4} \cmidrule(lr){5-7} \cmidrule(lr){8-10}
          ~ & CD${\downarrow}$ & F-Score${\uparrow}$ & IoU${\uparrow}$ & $RMSE_U{\downarrow}$ & $RMSE_O{\downarrow}$ & $\tau_o(\%){\uparrow}$  & $N_c{\downarrow}$ & $N_g{\downarrow}$ & $\tau_v(\%){\downarrow}$ \\
          \midrule 
          Ground Truth & 0.003584 & 1.000 & 1.000 & 6.017 & 7.980 & 99.77 & 17.8 & 32.4 & 0.0038\\
          \midrule
          \multicolumn{10}{@{}l@{}}{\textit{SS DiT/Shape DiT 512 + Shape VAE 512}} \\
          Pretrained pipeline & 0.031009 & 0.8937 & 0.5501 & 35.443 & 110.330 & 51.95 & 2803.8 & 100.7 & 11.3697 \\
          Finetuned pipeline & 0.045229 & 0.7616 & 0.3734 & 32.824 & 110.976 & 52.34 & 219.1 & 18.3 & 1.0238\\
          Finetuned pipeline + DE-Adapter v1 & 0.045209 & 0.7617 & 0.3734 & 32.799 & 55.146 & 87.55 & 219.5 & 19.4 & 1.0241\\
          Finetuned pipeline + DE-Adapter v2 & 0.045206 & 0.7619 & 0.3734 & 32.783 & 46.344 & 91.44 & 219.0 & 18.3 & 1.0237\\
          \midrule
          \multicolumn{10}{@{}l@{}}{\textit{SS DiT/Shape DiT 1024 + Shape VAE 512}} \\
          Pretrained pipeline & 0.030375 & 0.8959 & 0.5599 & 33.869 & 110.202 & 52.53 & 1724.9 & 1075.0 & 1.7782\\
          Finetuned pipeline & 0.040729 & 0.8253 & 0.4217 & 31.713 & 111.869 & 51.91 & 137.9 & 165.1 & 0.2129\\
          Finetuned pipeline + DE-Adapter v1 & 0.040751 & 0.8253 & 0.4216 & 31.746 & 80.133 & 75.12 & 138.5 & 167.3 & 0.2117\\
          Finetuned pipeline + DE-Adapter v2 & 0.040749 & 0.8253 & 0.4218 & 31.762 & 58.801 & 85.94 & 138.0 & 167.4 & 0.2143\\
          \bottomrule 
      \end{tabularx}
  \caption{Ablation study of DE-Adapter on shape generation under varying DiT resolutions and DE-Adapter architectures.}
  \label{tab:quant_ablation_fdg_d_shape_generation}
\end{table*}

\subsubsection{Ablation of DE-Adapter}
Tab.~\ref{tab:quant_ablation_fdg_d_shape_generation} reports the full \textbf{shape generation} pipeline (Image $\to$ DiTs $\to$ Shape VAE $\to$ FDG-D $\to$ Mesh). While DE-Adapter v1 (a simple MLPs architecture acting only on the final stage) improves normal quality over the finetuned baseline without hurting geometry or topology, its $\tau_o$ drops from $90.67\%$ in pure VAE reconstruction (Tab.~\ref{tab:quant_ablation_fdg_d_vae_reconstruction}) to $87.55\%$ in the full 512-resolution DiT generation pipeline. We attribute this performance gap to the noisy latents produced by DiT and the resulting domain shift from ground-truth latents. This issue is further compounded because v1 operates exclusively at the final high-resolution ($1\times$) stage, limiting its receptive field to local features. To address this, DE-Adapter v2 introduces a Feature Pyramid Network (FPN) architecture that fuses features across all five decoding stages ($16\times$ to $1\times$) by reusing the shape decoder's subdivision topology. This restores global context, successfully raising $\tau_o$ to $91.44\%$ and reducing $RMSE_O$ to $46.3^\circ$ while keeping the topology almost unchanged. Furthermore, our DE-Adapter design remains resolution-agnostic: the v2 adapter trained purely at 512 resolution can be directly applied to the $1024$-resolution generation pipeline (Tab.~\ref{tab:quant_ablation_fdg_d_shape_generation}, bottom) without any extra fine-tuning, still yielding a robust $\tau_o$ of $85.94\%$.

\subsubsection{Ablation of ROS-FT}
To comprehensively evaluate the contributions of individual training modules within our ROS-FT pipeline, we conduct an ablation study using the full 10K mixed 3D assets dataset (comprising open, closed, and hybrid surfaces). The quantitative results on the evaluation set are presented in Tab.~\ref{tab:quant_ablation_trained_modules}. Starting from the pretrained Trellis2 baseline, sequentially fine-tuning the Sparse Structure DiT (+A), Shape DiT (+B), and Shape Decoder (+C) on our dataset leads to progressive improvements in both Geometry Accuracy (e.g., F-Score rises from $0.8515$ to $0.8831$) and Topology Quality (e.g., non-manifold vertices $\tau_v$ drop from $1.86\%$ to $0.77\%$). However, fine-tuning these base modules alone (even when combined as +A \& B \& C) fails to resolve the inherent normal flipping issue, with the correct orientation rate ($\tau_o$) persistently remaining around $53\%$. It is only with the introduction of our final training stage---the FDG-D finetuned DE-Adapter (+D)---that the pipeline successfully overcomes this bottleneck. As shown in the final row (+ A \& B \& C \& D), the addition of the adapter dramatically boosts $\tau_o$ to $81.48\%$ and slashes the $RMSE_O$ error by nearly half (from $109.7$ to $63.348$), while strictly preserving the geometric and topological gains achieved in the earlier stages. This result firmly validates the effectiveness of our sequential training strategy and demonstrates that the DE-Adapter is a highly robust solution capable of restoring reliable face orientations across diverse surface types (open, closed, and hybrid) without compromising the priors from pre-training.

\begin{table*}[t]
  \centering
  \setlength{\tabcolsep}{4pt}
       \begin{tabularx}{\textwidth}{l
          >{\hsize=0.75\hsize\centering\arraybackslash}X
          >{\hsize=0.75\hsize\centering\arraybackslash}X
          >{\hsize=0.75\hsize\centering\arraybackslash}X
          >{\hsize=0.75\hsize\centering\arraybackslash}X
          >{\hsize=0.75\hsize\centering\arraybackslash}X
          >{\hsize=0.75\hsize\centering\arraybackslash}X
          >{\hsize=0.75\hsize\centering\arraybackslash}X
          >{\hsize=0.75\hsize\centering\arraybackslash}X
          >{\hsize=0.75\hsize\centering\arraybackslash}X}
          \toprule
          \multirow{2}{*}{Methods} & \multicolumn{3}{c}{Geometry Accuracy} & \multicolumn{3}{c}{Normal Quality} & \multicolumn{3}{c}{Topology Quality} \\
          \cmidrule(lr){2-4} \cmidrule(lr){5-7} \cmidrule(lr){8-10}
          ~ & CD${\downarrow}$ & F-Score${\uparrow}$ & IoU${\uparrow}$ & $RMSE_U{\downarrow}$ & $RMSE_O{\downarrow}$ & $\tau_o(\%){\uparrow}$  & $N_c{\downarrow}$ & $N_g{\downarrow}$ & $\tau_v(\%){\downarrow}$ \\
          \midrule
          Ground Truth & 0.004000 & 1.0000 & 1.0000 & 12.574 & 22.901 & 96.88 & 854.3 & 13.2 & 1.4328 \\
          Trellis2 & 0.042809 & 0.8515 & 0.5726 & 39.570 & 106.540 & 53.54 & 1564.9 & 1574.6 & 1.8692\\
          + SS DiT (A) & 0.036696 & 0.8835 & 0.6726 & 35.884 & 109.025 & 53.19 & 924.3 & 998.8 & 1.2513\\
          + Shape DiT (B) & 0.036668 & 0.8832 & 0.6729 & 34.728 & 109.644 & 53.11 & 574.2 & 736.2 & 0.8425\\
          + Shape Dec. (C) & 0.036676 & 0.8831 & 0.6729 & 34.597 & 109.776 & 53.06 & 518.4 & 727.1 & 0.7678\\
          + A \& B & 0.036683 & 0.8831 & 0.6729 & 34.599 & 109.719 & 53.06 & 520.0 & 727.7 & 0.7679\\
          + A \& C & 0.036682 & 0.8831 & 0.6728 & 34.627 & 109.771 & 53.02 & 519.8 & 727.5 & 0.7682\\
          + B \& C & 0.036682 & 0.8831 & 0.6728 & 34.589 & 109.709 & 53.07 & 519.3 & 728.9 & 0.7671\\
          + A \& B \& C & 0.036691 & 0.8831 & 0.6729 & 34.623 & 109.695 & 53.10 & 521.9 & 731.1 & 0.7702\\
          + A \& B \& C \& D & 0.036669 & 0.8831 & 0.6729 & 33.410 & 63.348 & 81.48 & 520.2 & 729.3 & 0.7681\\
          \bottomrule 
      \end{tabularx}
  \caption{Ablation study of ROS-FT on the mixed dataset (open, closed, and hybrid surfaces). \textit{Dec.} stands for decoder. \textit{+ A/B/C} indicates fine-tuning the corresponding module with its original loss. \textit{+ D} denotes the addition of the DE-Adapter, which is exclusively trained with the FDG-D representation in Stage 3.}
  \label{tab:quant_ablation_trained_modules}
\end{table*}

\begin{figure}[t]
  \centering
  \includegraphics[width=\linewidth]{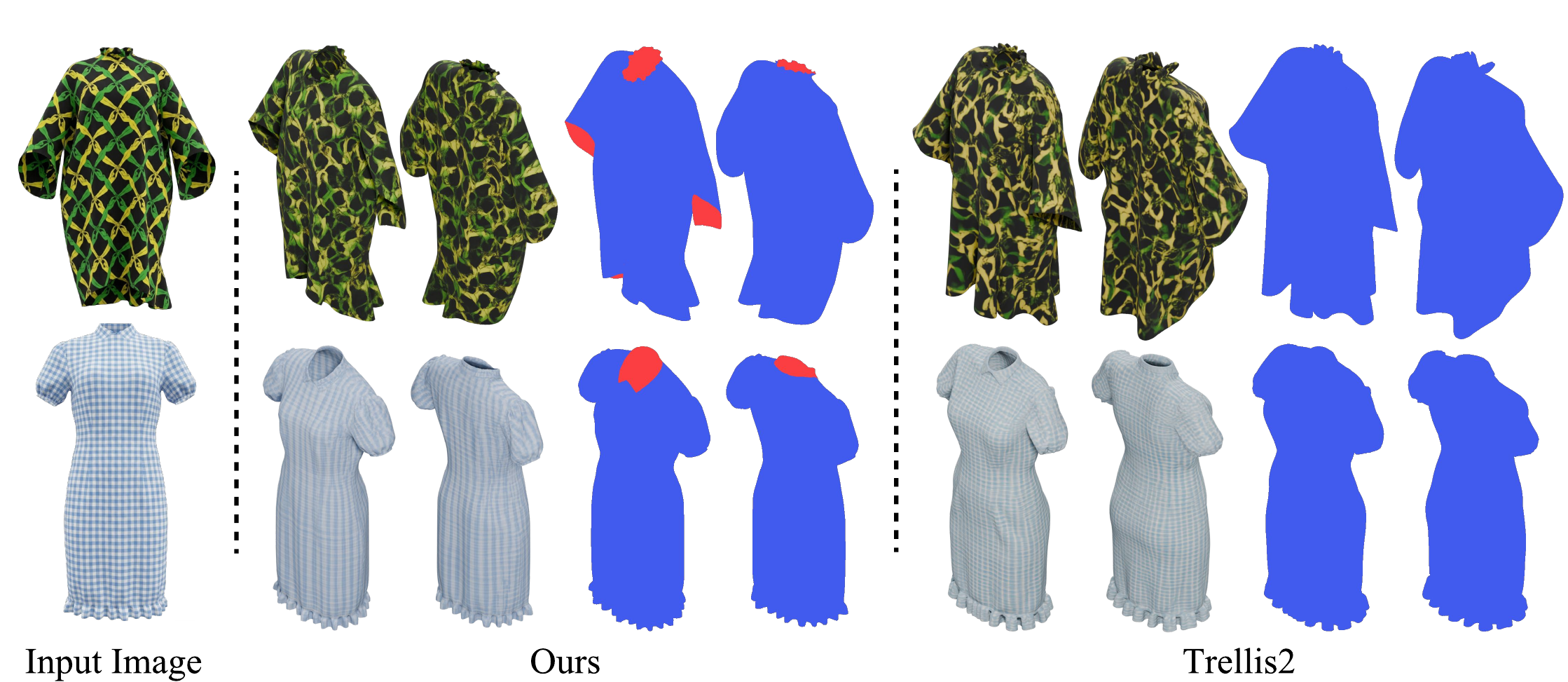}
  \vspace{-12pt}
  \caption{Our method is compatible with texturing pipeline of Trellis2.}
  \label{fig:6-texturing}
  \vspace{-8pt}
\end{figure}

\begin{figure}[t]
  \centering
  \includegraphics[width=\linewidth]{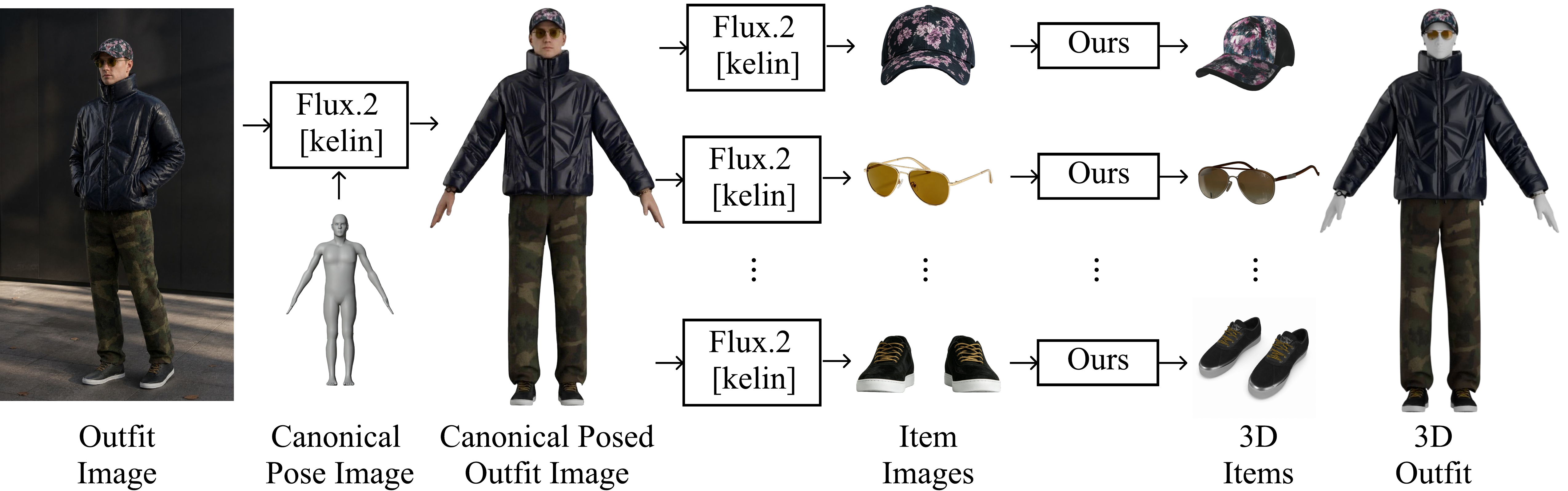}
  \vspace{-12pt}
  \caption{Multi-item 3D generation with our pipeline can generate open surfaces (pants) and closed surfaces (shoes) for a complex outfit.}
  \label{fig:6application}
  \vspace{-8pt}
\end{figure}

\section{Application and Discussion}

\begin{table}[t]
  \centering
  \setlength{\tabcolsep}{7pt}
  \renewcommand{\arraystretch}{1.15}
  \begin{tabular}{lrr}
    \hline
    Component & \#Params & Share (\%) \\
    \hline
    Sparse structure DiT & 1292.18M & 41.25 \\
    Sparse structure Decoder & 73.67M &  2.35 \\
    Shape DiT & 1292.25M & 41.25 \\
    Shape Decoder & 474.23M & 15.14 \\
    \hline
    Total & 3132.34M & 100.00 \\
    \textbf{DE-Adapter (ours)} & \textbf{36.38M} & \textbf{1.16} \\
    \hline
  \end{tabular}
  \caption{Parameter count breakdown of the 512 \emph{shape-only} inference pipeline (sparse structure DiT/Decoder + shape DiT/Decoder) and the relative size of the our proposed DE-Adapter.}
  \label{tab:adapter-param-ratio-shape-only-512}
\end{table}

\subsection{Texturing Pipeline Compatibility}
The original textured mesh generation in Trellis2 necessitates a post-processing remeshing stage, where the decoded open surface is converted into a narrow-band Unsigned Distance Function (UDF) and subsequently reconstructed via dual contouring. This procedure ensures consistent face orientations for texture baking, but it introduces topological distortions by closing open boundaries. This effect is evidenced in the rightmost columns of Fig.~\ref{fig:6-texturing}, where only the front-facing (blue) orientations. In contrast, our pipeline generates high-fidelity, globally consistent normals directly, allowing us to bypass this destructive remeshing step. By eliminating the need for UDF-based reconstruction, we facilitate a seamless transition to downstream texturing while strictly preserving the intended open-surface topology of the generated 3D assets.

\subsection{Multi-item Generation Pipeline}
Because our AnySurf pipeline is inherently designed as a unified, one-for-any surface generator, it can seamlessly harness the advanced image generation models like Flux.2 [klein]~\cite{flux2klein} and ChatGPT Images 2.0~\cite{gptimage2}. As demonstrated in Fig.~\ref{fig:6application}, this synergy allows our method to effortlessly and reliably generate open-surface garments (e.g., pants) alongside their matching closed-surface accessories (e.g., shoes) within one complex outfit. More results are available in our demo video.

\subsection{Parameter Efficiency}
\label{sec:param_efficiency}
To demonstrate the lightweight nature of our approach, we provide a parameter count breakdown for the 512-resolution \emph{shape-only} inference pipeline (comprising the sparse structure flow/decoder, shape flow, and shape decoder). As shown in Tab.~\ref{tab:adapter-param-ratio-shape-only-512}, our introduced ROS adapter adds only a marginal parameter overhead relative to the frozen base model, ensuring efficient fine-tuning without compromising the generative capabilities of the pre-trained backbone. 

\subsection{Further Improvements}
While our proposed method achieves significant improvements, its performance could be further elevated through two primary scaling directions. First, \textit{Data Scaling}: we finetuned the base model on approximately 10K assets, accounting for only $\sim1\%$ of the 970K+ pre-training data used by the base model. Scaling up the high-quality fine-tuning dataset to match the pre-training scale would undoubtedly enhance generalization across more diverse topologies. Second, \textit{Architecture Scaling}: our current DE-Adapter introduces merely $\sim1\%$ additional parameters relative to base model. Expanding the parameter size could capture more complex geometric contexts. Ultimately, the most optimal solution might involve designing a similar orientation-aware module natively within the base SC-VAE and incorporating face orientation learning directly into the large-scale pre-training phase, rather than relying solely on post-training.

\begin{figure}[t]
	\centering
	\includegraphics[width=\linewidth]{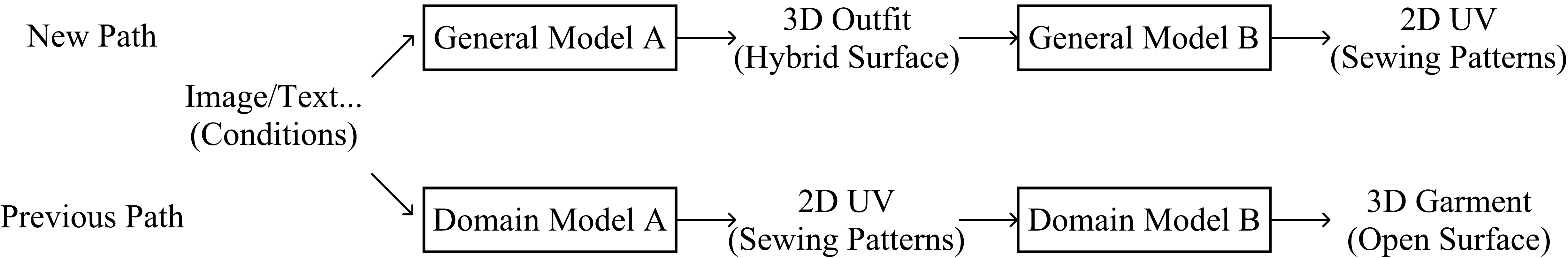}
	\caption{A new unified pathway for 3D fashion assets generation.}
	\label{fig:future_directions}
\end{figure}

\subsection{Future Directions}
Our work establishes a new pathway for 3D fashion modeling by transitioning from domain-specific pattern-stitching to a universal 3D topological generation paradigm. As illustrated in Fig.~\ref{fig:future_directions}, this new pathway overcomes the fundamental limitations of the traditional 2D pattern-stitching approach by directly generating open, closed, and hybrid topologies in a unified 3D space. While our proposed method serves as a crucial first step by addressing the normal limitations of general 3D generators (General Model A in Fig.~\ref{fig:future_directions}), achieving fully controllable, production-ready 3D fashion assets generation remains an open challenge. At the next step, combining our robust 3D outfits with the UV and seam generation techniques—such as PartUV~\cite{wang2025partuv} and MeshTailor~\cite{ma2026meshtailorcuttingseamsgenerative} (General Model B in Fig.~\ref{fig:future_directions})—will pave the way toward a fully automated and universal 3D fashion assets generation pipeline.

\section{Conclusion}
In this work, we present AnySurf, a unified 3D generation pipeline capable of producing open, hybrid, and closed surfaces with accurate face orientations. To overcome the inherent limitations of original FDG representation, we introduce FDG-D (Flexible Dual Grid with Directed Edge), a minimal yet effective 3D representation that preserves critical normal information across diverse topologies. Furthermore, we propose ROS-FT, an efficient post-training strategy utilizing a lightweight DE-Adapter (adding only 1.16\% parameters) to significantly enhance normal quality without compromising the base model's original generative capabilities. To validate our approach and facilitate future research, we construct Outfit3D, an industry-level hybrid dataset featuring high-precision garments and accessories with rich textures and UV information. Extensive experiments demonstrate that AnySurf significantly outperforms existing baselines across diverse surface topologies, evolving 3D fashion generation from an isolated \textit{domain-specific} task into an integral component of the \textit{general} 3D generative landscape.


\bibliographystyle{ACM-Reference-Format}
\bibliography{sample-base}

\end{document}